\documentclass[prd,preprint,floatfix,nofootinbib,preprintnumbers,superscriptaddress,showkeys]{revtex4} 
\pdfoutput=1
\usepackage[caption=false]{subfig}
\usepackage{amsmath}
\usepackage{amsfonts}
\usepackage{amssymb}
\usepackage{float}
\usepackage{color}
\usepackage{graphicx}
\usepackage{graphics}
\usepackage{hyperref}
\usepackage{adjustbox,array}
\usepackage{enumitem} 
\hypersetup{}
\usepackage[utf8x]{inputenc} 
\usepackage{multirow}
\usepackage{booktabs}
\usepackage{mathrsfs}
\usepackage{diagbox}
\usepackage[english]{babel}
\usepackage[autostyle]{csquotes}

\graphicspath{{./figures/}}

\definecolor{rosso}{cmyk}{0,1,1,0.4}
\definecolor{rossos}{cmyk}{0,1,1,0.55}
\definecolor{rossoc}{cmyk}{0,1,1,0.2}
\definecolor{blu}{cmyk}{1,1,0,0.3}
\definecolor{blus}{cmyk}{1,1,0,0.6}
\definecolor{bluc}{cmyk}{1,1,0,0.1}
\definecolor{verde}{cmyk}{0.92,0,0.59,0.25}
\definecolor{verdec}{cmyk}{0.92,0,0.59,0.15}
\definecolor{verdes}{cmyk}{0.92,0,0.59,0.4}

\AtBeginDocument{
\heavyrulewidth=.08em
\lightrulewidth=.05em
\cmidrulewidth=.03em
\belowrulesep=.65ex
\belowbottomsep=0pt
\aboverulesep=.4ex
\abovetopsep=0pt
\cmidrulesep=\doublerulesep
\cmidrulekern=.5em
\defaultaddspace=.5em
}

\hypersetup{
 colorlinks=true,
 linkcolor=verdec,
 urlcolor=verdec,
 citecolor=verdec
}

\newcommand{ \eq}[1]{Eq.~(\ref{#1})}

\newcommand{\gsim}{\gtrsim}
\newcommand{\lsim}{\lesssim}

\newcommand{\lf}{\left(}
\newcommand{\ri}{\right)}

\newcommand{\nn}{\nonumber}

\newcommand{\sqt}{\sqrt{2}}

\renewcommand{\lg}{\mathscr{L}} 

\newcommand{\br}{{\rm Br}}
\newcommand{\hc}{{\rm H.c.}}

\newcommand{\gev}{{\;{\rm GeV}}}
\newcommand{\tev}{{\;{\rm TeV}}}

\newcommand{\beq}{\begin{equation}}
\newcommand{\eeq}{\end{equation}}
\newcommand{\bea}{\begin{eqnarray}}
\newcommand{\eea}{\end{eqnarray}}
\newcommand{\barr}{\begin{array}}
\newcommand{\earr}{\end{array}}
\newcommand{\bc}{\begin{center}}
\newcommand{\ec}{\end{center}}
\newcommand{\bit}{\begin{itemize}}
\newcommand{\eit}{\end{itemize}}
\newcommand{\ben}{\begin{enumerate}}
\newcommand{\een}{\end{enumerate}}

\newcommand{\al}{\alpha}
\newcommand{\bt}{\beta}

\newcommand{\Dt}{\Delta}

\newcommand{\sg}{\sigma}
\newcommand{\es}{\epsilon}

\newcommand{\gm}{\gamma}

\newcommand{\lm}{\lambda}

\newcommand{\lmc}{\Lambda_{\rm c}}

\newcommand{\hsm}{{h_{\rm SM}}}
\newcommand{\ch}{H^\pm}

\newcommand{\mh}{m_{h}}
\newcommand{\mch}{M_{H^\pm}}
\newcommand{\mhh}{M_{H}}
\newcommand{\ma}{M_{A}}
\newcommand{\dmh}{\Delta m_h}
\newcommand{\dmhh}{\Delta M_{H}}
\newcommand{\dma}{\Delta M_{A}}
\newcommand{\mhsm}{m_{125}}

\newcommand{\ca}{c_\alpha}
\newcommand{\sa}{s_\alpha}

\newcommand{\sta}{s_{2\alpha}}

\newcommand{\tb}{t_\beta}

\newcommand{\cb}{c_\beta}
\renewcommand{\sb}{s_\beta}
\newcommand{\stb}{s_{2\beta}}

\newcommand{\cba}{c_{\beta-\alpha}}
\newcommand{\sba}{s_{\beta-\alpha}}

\newcommand{\ee}      {{e^+ e^-}}

\newcommand{\ttau}      {{\tau^+\tau^-}} 

\definecolor{mint}{rgb}{0.24, 0.71, 0.54}

\begin{document}

\title{\color{verdes} Status of the two-Higgs-doublet model\\ in light of the CDF $m_W$ measurement}
\author{Soojin Lee}
\email{soojinlee957@gmail.com}
\address{Department of Physics, Konkuk University, Seoul 05029, Republic of Korea}
\author{ Kingman Cheung}
\email{cheung@phys.nthu.edu.tw}
\address{Department of Physics, Konkuk University, Seoul 05029, Republic of Korea}
\address{Department of Physics, National Tsing Hwa University, Hsinchu 300, Taiwan}
\address{Center for Theory and Computation, National Tsing Hua University,
  Hsinchu 300, Taiwan}
\author{Jinheung Kim}
\email{jinheung.kim1216@gmail.com}
\address{Department of Physics, Konkuk University, Seoul 05029, Republic of Korea}
\author{Chih-Ting Lu}
\email{timluyu@gmail.com}
\address{Department of Physics and Institute of Theoretical Physics, Nanjing Normal University, Nanjing, 210023, China}
\author{Jeonghyeon Song}
\email{jhsong@konkuk.ac.kr}
\address{Department of Physics, Konkuk University, Seoul 05029, Republic of Korea}

\begin{abstract}
The most recent $W$-boson mass measurement by the CDF collaboration with a substantially reduced uncertainty indicates a significant deviation from the standard model prediction, as large as $7\sigma$ if taken literally.  Then the Peskin-Takeuchi parameters of $S$ and $T$ shift to larger values, which has profound consequences in searching for physics beyond the SM. In the framework of two-Higgs-doublet models, we study the effect of the new $W$-boson mass measurement on the parameter space. Combined with other constraints including theoretical requirements, flavor-changing neutral currents in $B$ physics, the cutoff scale above 1 TeV, Higgs precision data, and direct collider search limits from the LEP, Tevatron, and LHC experiments, we find upper bounds on the masses of the heavy Higgs bosons: $M_{H, A, H^\pm} \lesssim 1.1$ TeV in type I, II, X, and Y for the normal Higgs scenario; $M_{H^\pm} \lesssim 450 $ GeV and   $M_{A} \lesssim 420 $ GeV in type I and X for the inverted scenario where the heavier $CP$-even Higgs bosons is the observed one. Another important finding is that type II and type Y in the inverted scenario are completely excluded. Such unprecedented findings imply that the upcoming LHC run can readily close out a large portion of the still-available parameter space.
\end{abstract}

\vspace{1cm}

\maketitle
\tableofcontents

\section{Introduction}

The standard model (SM) of electroweak theory with $ SU(2)\times U(1) $ gauge symmetry is highly successful in explaining almost all the measurements in particle physics experiments. 
Nevertheless,
we have not given up on new physics beyond the SM (BSM)
to answer the outstanding questions in particle physics, such as the neutrino mass and mixing, 
matter-antimatter asymmetry in the Universe, and dark matter.
A natural approach is to presume that a BSM theory appears at a high energy scale
while the SM is a good theory at a low energy scale. 
Hence, if any experiment at the energy scale below 1 TeV observes an anomaly,
it would shake the foundation on the SM and indicate the advent of a new era in particle physics.

Very recently, the CDF collaboration at Fermilab reported the most precise $W$-boson mass measurement,
$m^{\text{CDF}}_W = 80.4335\pm 0.0094$ GeV~\cite{CDF:2022hxs}.
The total uncertainty is less than $10$ MeV  and 
the central value is about $76.5$ MeV larger than the SM prediction, $m^{\text{SM}}_W = 80.357\pm 0.006$ GeV~\cite{ParticleDataGroup:2020ssz}.
It has about $7\sigma$ standard deviation from the SM value. 
Before the CDF run-II result,
the world average of $m_W$ measurements had just $1.8\,\sigma$ standard deviation from $m^{\text{SM}}_W$~\cite{ParticleDataGroup:2020ssz}. 
Even though more careful cross-checks of the systematic uncertainties between CDF run-II analysis and 
other $W$-boson mass measurements at the LEP~\cite{ALEPH:2013dgf}, LHCb~\cite{LHCb:2021bjt}, ATLAS~\cite{ATLAS:2017rzl}, and D0~\cite{D0:2012kms} will eventually be made, 
this new measured $m_W$ value urgently calls for an explanation 
from new physics models. 

An efficient parametrization to quantify the validity of the SM and
to point in the direction of new physics is
a set of the Peskin-Takeuchi oblique parameters of $S$, $T$, and $U$~\cite{Peskin:1990zt,Marciano:1990dp,Kennedy:1990ib}
in the global fit to the electroweak precision data (EWPD)~\cite{Haller:2018nnx,deBlas:2021wap}. 
According to Ref.~\cite{Lu:2022bgw,Strumia:2022qkt,deBlas:2022hdk,Fan:2022yly,Paul:2022dds,Gu:2022htv,Asadi:2022xiy,Endo:2022kiw,Balkin:2022glu,Carpenter:2022oyg,Du:2022fqv}, 
the CDF $m_W$ yields significant deviations of the oblique parameters from the SM predictions.
If all of three can vary, 
the new fits show that $S$ and $T$ can keep as before, 
but the $U$ increases substantially such that $S=0.06\pm 0.10$, $T=0.11\pm 0.12$, 
and $U=0.13\pm 0.09$~\cite{Lu:2022bgw}. 
Here, we take the definition of the oblique parameters which vanish in the SM~\cite{ParticleDataGroup:2020ssz}. 
However, the contributions to $U$ can only appear in a dimension-eight operator, so most new physics models have tiny contributions to $U$. 
Therefore, setting $U=0$ while varying $S$ and $T$ is usually adopted,
which results in both $S$ and $T$ moving to large and positive values: $S=0.15\pm 0.08$ and $T=0.27\pm 0.06$~\cite{Lu:2022bgw}. 
Based on these changes of $S$ and $T$, some new physics models including two-Higgs-doublet model (2HDM) and its extensions~\cite{Fan:2022dck,Zhu:2022tpr,Lu:2022bgw,Zhu:2022scj,Song:2022xts,Bahl:2022xzi,Heo:2022dey,Babu:2022pdn,Biekotter:2022abc,Ahn:2022xeq,Han:2022juu,Arcadi:2022dmt,Ghorbani:2022vtv,Broggio:2014mna,Kim:2022hvh}, the Higgs triplet model~\cite{Cheng:2022jyi,Du:2022brr,Kanemura:2022ahw,Mondal:2022xdy,Borah:2022obi}, supersymmetry~\cite{Yang:2022gvz,Du:2022pbp,Tang:2022pxh,Athron:2022isz,Zheng:2022irz,Ghoshal:2022vzo}, leptoquarks~\cite{Athron:2022qpo,Cheung:2022zsb,Bhaskar:2022vgk}, seesaw mechanisms of neutrino mass~\cite{Blennow:2022yfm,Arias-Aragon:2022ats,Liu:2022jdq,Chowdhury:2022moc,Popov:2022ldh,Batra:2022pej}, vector-like leptons or vector-like quarks~\cite{Lee:2022nqz,Kawamura:2022uft,Crivellin:2022fdf,Nagao:2022oin,Cao:2022mif},
the standard model effective field theory (SMEFT)~\cite{deBlas:2022hdk,Fan:2022yly,Bagnaschi:2022whn,Paul:2022dds,Gu:2022htv,DiLuzio:2022xns,Endo:2022kiw,Balkin:2022glu,Cirigliano:2022qdm}, and others~\cite{Yuan:2022cpw,Strumia:2022qkt,Cacciapaglia:2022xih,Sakurai:2022hwh,Heckman:2022the,Krasnikov:2022xsi,Peli:2022ybi,Perez:2022uil,Wilson:2022gma,Zhang:2022nnh,Carpenter:2022oyg,Du:2022fqv} are proposed to explain the $W$-boson mass anomaly. In particular, some of them also try to explain the long-standing anomaly 
in the muon anomalous magnetic moment measurement~\cite{Athron:2022qpo,Du:2022pbp,Tang:2022pxh,Lee:2022nqz,Han:2022juu,Kawamura:2022uft,Cheung:2022zsb,Nagao:2022oin,Chowdhury:2022moc,Arcadi:2022dmt,Bhaskar:2022vgk,Kim:2022hvh}, 
which the Fermilab has recently confirmed~\cite{Muong-2:2021ojo}.

In this work, we pursue a comprehensive study of the 2HDM in light of the new CDF $m_W$ measurement.
We study not only 
four famous tree-level flavor-conserved types (type I, type II, type X, and type Y)
but also two Higgs scenarios for the observed Higgs boson, 
the normal scenario (NS) and inverted scenario (IS). 
We impose all the theoretical and experimental constraints, including the stability of the scalar potential,
the unitarity of the scalar-scalar scatterings,
the EWPD, the Higgs precision data, and the direct search bounds at the LEP, Tevatron, and LHC. 
We compare the results before and after the new CDF $m_W$ measurement.
In addition, we study the evolutions of the model parameters via renormalization group equation (RGE),
and demand the stability of the scalar potential up to 1 TeV.
A particular focus is on a comparative study 
to see the differences in the viable parameter space, according to the type, Higgs scenario,
and $m_W$.
One of the most salient features when we take $m_W^{\rm CDF}$ is
that the aforementioned constraints 
put the \emph{upper} bounds on the masses of new Higgs bosons,
about 1.1 TeV in the NS and about 450 GeV in the IS.
Then type II and type Y in the IS face a conflict with the lower bound on $\mch\gsim 580\gev$ from $b\to s\gm$~\cite{Arbey:2017gmh,Misiak:2017bgg}.
Consequently, type II and type Y in the IS are excluded.
In addition,
the results of scanning the entire parameter space
without any conditions on the masses and couplings
give apparent signals for the future collider phenomenologies:
(i) type I has the most surviving parameter points for both NS and IS;
(ii) the light charged Higgs boson at a mass below the top quark mass is viable in type I and type X;
(iii) light neutral Higgs bosons, \textit{CP}-even and  \textit{CP}-odd, are still allowed for type I and type X.

The rest of this paper is arranged as follows. We briefly review the 2HDM in Sec.~\ref{sec:review}. 
The parameter scanning strategies are outlined in Sec.~\ref{sec:scan}.
The allowed ranges of the masses, $\tan\beta$, and $\sin (\bt-\al)$ are also shown. 
We then discuss the characteristic features of the NS and IS 
 in Sec.~\ref{sec:NS} and Sec.~\ref{sec:IS}, respectively. Finally, we conclude in Sec.~\ref{sec:conclusions}.

\section{Review of 2HDM}
\label{sec:review}

In the 2HDM,
there exist two complex $SU(2)_L$ Higgs doublet fields, $\Phi_1$ and $\Phi_2$~\cite{Branco:2011iw}:
\bea
\label{eq:phi:fields}
\Phi_i = \left( \begin{array}{c} w_i^+ \\[3pt]
\dfrac{v_i +  h_i + i \eta_i }{ \sqrt{2}}
\end{array} \right), \quad i=1,2,
\eea
where $v_{1}$ and $v_2$ are the nonzero vacuum expectation values of $\Phi_1$ and $\Phi_2$, respectively.
The electroweak symmetry is broken by $v =\sqrt{v_1^2+v_2^2}=246\gev $.
We define the ratio of two vacuum expectation values to be $\tan \beta =v_2/v_1$.
For simplicity, we use the notation of
  $s_x=\sin x$, $c_x = \cos x$, and $t_x = \tan x$ in what follows.
  
We impose a discrete $Z_2$ symmetry,
under which $\Phi_1 \to \Phi_1$
and $\Phi_2 \to -\Phi_2$, to avoid the flavor-changing-neutral-current (FCNC)
at tree level~\cite{Glashow:1976nt,Paschos:1976ay}.
The scalar potential with $CP$ invariance  and softly broken $Z_2$  is
\bea
\label{eq:VH}
V = && m^2 _{11} \Phi^\dagger _1 \Phi_1 + m^2 _{22} \Phi^\dagger _2 \Phi_2
-m^2 _{12} ( \Phi^\dagger _1 \Phi_2 + \hc) \\ \nn
&& + \frac{1}{2}\lambda_1 (\Phi^\dagger _1 \Phi_1)^2
+ \frac{1}{2}\lambda_2 (\Phi^\dagger _2 \Phi_2 )^2
+ \lambda_3 (\Phi^\dagger _1 \Phi_1) (\Phi^\dagger _2 \Phi_2)
+ \lambda_4 (\Phi^\dagger_1 \Phi_2 ) (\Phi^\dagger _2 \Phi_1) \\ \nn
&& + \frac{1}{2} \lambda_5
\left[
(\Phi^\dagger _1 \Phi_2 )^2 +  \hc
\right],
\eea
where the $m^2 _{12}$ term softly breaks the $Z_2$ parity.
The model has five physical Higgs bosons, the lighter \textit{CP}-even scalar $h$,
the heavier \textit{CP}-even scalar $H$, the \textit{CP}-odd pseudoscalar $A$,
and a pair of charged Higgs bosons $H^\pm$.
The weak eigenstates in Eq.~(\ref{eq:phi:fields}) are linear combinations of physical Higgs bosons 
through two mixing angles, $\al$ and $\bt$: the expressions are referred to Ref.~\cite{Song:2019aav}.
An important relationship is the SM Higgs boson $\hsm$ with $h$ and $H$:
\bea
\label{eq:hsm}
\hsm = \sba h + \cba H.
\eea

In the 2HDM, the observed Higgs boson at a mass of 125 GeV can be either $h$ or $H$,
which is called the normal scenario (NS) and the inverted scenario (IS)~\cite{Chang:2015goa,Jueid:2021avn}, respectively:
\begin{align}
\label{eq:NS:IS}
\hbox{NS: } &~ \mh=\mhsm;
\\ \nn
\hbox{IS: } &~ \mhh=\mhsm,
\end{align}
where $\mhsm=125\gev$.
A popular way to accommodate the SM-like Higgs boson
is the Higgs alignment limit where $\hsm=h$ (or $\cba=0$) in the NS and $\hsm=H $ (or $\sba=0$) in the IS.
Then the phenomenology of the BSM Higgs bosons is simplified such that 
$H \to WW/ZZ$, $A \to Z \hsm$, and $\ch\to W^{\pm (*)}\hsm$ are prohibited at tree level.
However, the assumption may interfere with observing new scalar bosons at the LHC.
Therefore, we do not impose any conditions on the masses and couplings when performing the random scan.
Only the theoretical and experimental constraints will restrict the parameter space.

We take six free parameters of
\bea
\label{eq:model:parameters}
\left\{\mh,\quad \mch, \quad\mhh,\quad \ma,\quad m_{12}^2,\quad \tb,\quad\sba \right\}.
\eea
The range of $(\bt-\al)$ is set to be $[-\pi/2,\pi/2]$, 
as in the public codes of \textsc{2HDMC}~\cite{Eriksson:2009ws},
\textsc{HiggsSignals}~\cite{Bechtle:2020uwn}, and \textsc{HiggsBounds}~\cite{Bechtle:2020pkv}.
The quartic couplings are~\cite{Das:2015mwa}
\bea
\label{eq:quartic}
\lm_1  &=& \frac{1}{v^2 \cb^2}
\left[
\ca^2 \mhh^2 + \sa^2 m_h^2 - \sb^2 M^2
\right], \\[3pt] \nn
\lm_2 &=& \frac{1}{v^2 \sb^2}
\left[
\sa^2 \mhh^2+\ca^2 m_h^2 - \cb^2 M^2
\right], \\[3pt] \nn
\lm_3 &=& 
\frac{1}{v^2}
\left[
2 \mch^2 + \frac{\sta}{\stb} (\mhh^2-m_h^2) - M^2
\right], \\[3pt] \nn
\lm_4 &=& 
\frac{1}{v^2}
\left[
\ma^2- 2 \mch^2 + M^2
\right], \\[3pt] \nn
\lm_5 &=&
\frac{1}{v^2}
\left[
M^2-\ma^2
\right], 
\eea
where $M^2 = m_{12}^2/(\sb\cb)$.

\begin{table}[t]
\begin{center}
 {\renewcommand{\arraystretch}{1.2} 
\begin{tabular}{|c||c|c|c||c|c|c||c|c|c|}
\hline
& ~~$\xi^h_u$~~ & ~~$\xi^h_d$~~ & ~~$\xi^h_\ell$~~
& ~~$\xi^H_u$~~ & ~~$\xi^H_d$~~ & ~~$\xi^H_\ell$~~
& ~~$\xi^A_u$~~ & ~~$\xi^A_d$~~ & ~~$\xi^A_\ell$~~ \\ \hline
~~~type I~~~
& $\frac{\ca}{\sb}$ & $\frac{\ca}{\sb} $ & $\frac{\ca}{\sb} $
& $\frac{\sa}{\sb} $ & $\frac{\sa}{\sb} $ & $\frac{\sa}{\sb} $
& $\frac{1}{\tb} $ & $-\frac{1}{\tb}$ & $-\frac{1}{\tb}$ \\
type II
& $\frac{\ca}{\sb} $ & $-\frac{\sa}{\cb}$ & $-\frac{\sa}{\cb}$
& $\frac{\sa}{\sb}$ & $\frac{\ca}{\cb} $ & $\frac{\ca}{\cb}$
& $\frac{1}{\tb} $ & $\tb$ & $\tb$ \\
type X
& $\frac{\ca}{\sb}$ & $\frac{\ca}{\sb}$ & $-\frac{\sa}{\cb} $
& $\frac{\sa}{\sb} $ & $\frac{\sa}{\sb}$ & $\frac{\ca}{\cb}$
& $\frac{1}{\tb} $ & $-\frac{1}{\tb}$ & $\tb$ \\
type Y
& $\frac{\ca}{\sb} $ & $-\frac{\sa}{\cb}$ & $ \frac{\ca}{\sb} $
& $\frac{\sa}{\sb} $ & $\frac{\ca}{\cb}$ & $\frac{\sa}{\sb}$
& $\frac{1}{\tb} $ & $\tb$ & $-\frac{1}{\tb}$ \\
\hline
\end{tabular}
}
\end{center}
\caption{The Yukawa coupling modifiers in four types of the 2HDM. }\label{tab:Yukawa}
\end{table}

According to the $Z_2$ parity of the fermion singlets,
there are four types in the 2HDM,
type I, type II, type X, and type Y,
which have different Yukawa couplings of the SM fermions.
We parametrize the Yukawa Lagrangian as
\bea
\label{eq:Lg:Yukawa}
\lg_{\rm Yuk} &=&
- \sum_f 
\lf 
\frac{m_f}{v} \xi^h_f \bar{f} f h + \frac{m_f}{v} \xi^H_f \bar{f} f H
-i \frac{m_f}{v} \xi^A_f \bar{f} \gm_5 f A
\ri
\\ \nn &&
- 
\left\{
\dfrac{\sqrt2V_{ud}}{v } H^+  \overline{u}
\left(m_u \xi^A_u \text{P}_L +  m_d \xi^A_d \text{P}_R\right)d 
+\dfrac{\sqt m_\ell}{v}H^+ \xi^A_\ell \overline{\nu}_L\ell_R^{}
+\hc
\right\},
\eea
where $\xi^{h,H,A}_f$ are presented in Table \ref{tab:Yukawa}.
To specify the eight cases (four types in the NS and four types in the IS), 
we shall sometimes use a simplified name: for instance, NS-I denotes type I in the NS.

Now let us get into the comparative study in the 2HDM before and after the CDF $m_W$ measurement, 
denoted by the \enquote{PDG} and \enquote{CDF} cases respectively.
The Peskin-Takeuchi oblique parameters with $U=0$ in two cases are
\begin{align}
	\label{eq:STU:PDG}
	S_{\rm PDG} &= 0.05 \pm 0.08,
	\quad
	T_{\rm PDG} = 0.09 \pm 0.07, \quad
	\rho_{\rm PDG} = 0.92, 
	\\ \label{eq:STU:CDF}
	 S_{\rm CDF}&= 0.15 \pm 0.08,
	\quad
	T_{\rm CDF} = 0.27 \pm 0.06, \quad
	\rho_{\rm CDF} = 0.93, 
	\end{align}
	where $\rho$ is the correlation between $S$ and $T$.

\begin{figure}[t] \centering
\begin{center}
\includegraphics[width=0.42\textwidth]{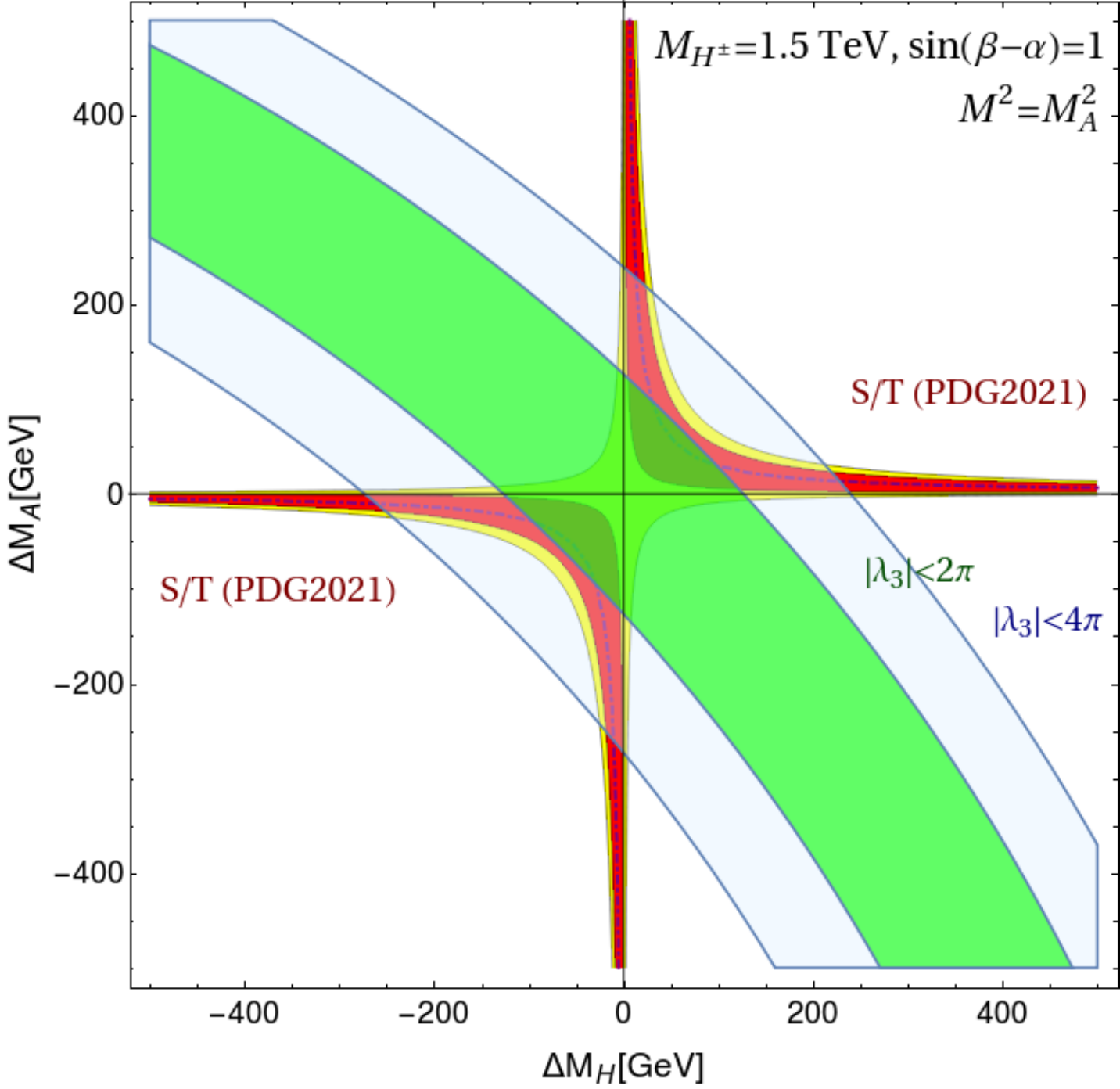}~~~~
\includegraphics[width=0.42\textwidth]{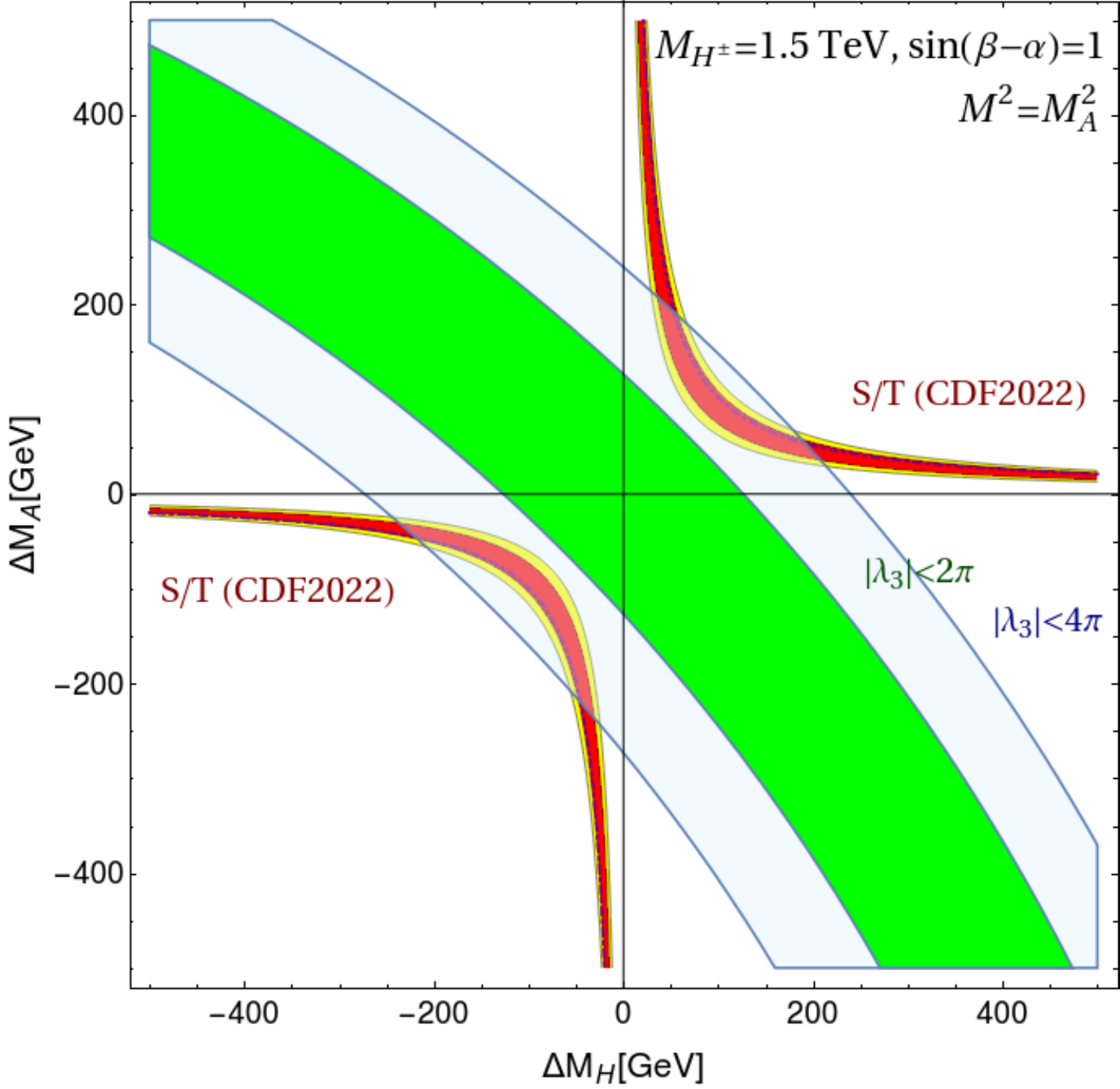}
\end{center}
\caption{\label{fig:ST:lm3:dMH:dMA}
Allowed regions of $(\dmhh,\dma)$ by the oblique parameters of $S$ and $T$ (red at $1\sg$ and yellow in $2\sg$), 
$|\lm_3|<4 \pi$ (light blue),
and $|\lm_3|<2\pi$ (green),
where $\Dt m \equiv m-\mch$.
The left (right) panel shows the PDG (CDF) results.
We set $\mch=1.5\tev$ and $M^2=\ma^2$ in the normal scenario with $\sba=1$.
}
\end{figure}

The biggest difference between the PDG and CDF cases is that the CDF $m_W$ does not allow new Higgs bosons heavier than about $1.1\tev$, while the PDG allows, which will be explicitly shown in the next section.
To reveal the origin of this key feature,
let us consider heavy charged Higgs bosons with $\mch=1.5\tev$.
For simplicity, we concentrate on the NS in the Higgs alignment limit ($\sba=1$),
where the quartic couplings are
\bea
\label{eq:quartic:NS:aligned}
\lm_1^{\rm NS-Al}  &=& \frac{1}{v^2 }
\left[
\tb^2 (\mhh^2 - M^2) -  m_h^2 
\right], \\ \nn
\lm_2^{\rm NS-Al} &=& \frac{1}{v^2 }
\left[ \mh^2 + \frac{1}{\tb^2} \lf \mhh^2-M^2\ri
\right], \\ \nn
\lm_3^{\rm NS-Al} &=& 
\frac{1}{v^2}
\left[ \mh^2 +2 \mch^2 -\mhh^2-M^2
\right].
\eea
$\lm_4^{\rm NS-Al}$ and $\lm_5^{\rm NS-Al}$ are the same as  in \eq{eq:quartic}.
It is well known in the literature~\cite{Chakrabarty:2014aya,Chowdhury:2015yja,Bagnaschi:2015pwa,Ferreira:2015rha,Das:2015mwa,Chakrabarty:2016smc,Cacchio:2016qyh,Chowdhury:2017aav,Chakrabarty:2017qkh,Basler:2017nzu,Branchina:2018qlf,Dey:2021pyn,Kim:2022nmm} that
if any quartic coupling at the electroweak scale is not small enough,
its magnitude grows rapidly as the energy scale increases, ending up with the breaking of the stability of the scalar potential.
For illustration purposes, 
let us select $\lm_3$ among five quartic couplings 
because it is independent of $\tb$ but sensitive to three new masses.\footnote{In the next section, we will
perform the complete RGE analysis for the gauge, Yukawa, and quartic couplings.
}
Figure \ref{fig:ST:lm3:dMH:dMA} shows $(\dmhh,\dma)$ with $M^2 = \ma^2$,
allowed by $S$ and $T$ at $1\sg$ level (red), at $2\sg$ level (yellow), $|\lm_3|<4 \pi$ (light blue),
and $|\lm_3|<2 \pi$ (green).
The PDG result is in the left panel, and the CDF result is in the right panel.
The condition of $|\lm_3|<4\pi$ is for the perturbativity of the quartic coupling,
and $|\lm_3|< 2\pi$ is for $\lmc>1\tev$:
the bound of $2\pi$ is chosen
because it is the maximum of $|\lm_3|$ allowed by $\lmc>1\tev$ in the full RGE analysis.

Figure \ref{fig:ST:lm3:dMH:dMA} clearly shows the difference between the PDG and CDF cases.
First, the oblique parameters allow $\mch=\mhh=\ma$
in the PDG case, but not in the CDF case.
Notice that heavy masses of new Higgs bosons are still consistent with $T_{\rm CDF}$ in \eq{eq:STU:CDF}:
for example, very heavy $\ma$ is feasible if $\dmhh\simeq 0$.
When applying the perturbativity of $|\lm_3|<4\pi$  (light blue),
however,
large mass gaps of $\dmhh$ and $\dma$ are forbidden in both cases.
When narrowing the range further into $|\lm_3|<2\pi$ for $\lmc>1\tev$,
there is no overlap in the CDF case.
Thus heavy Higgs bosons cannot simultaneously satisfy
$T_{\rm CDF}$  and $\lmc>1\tev$.
For lighter $\mch$, however,
the CDF case also permits an overlap because of \eq{eq:quartic:NS:aligned}.

\section{Scanning strategies and the results}
\label{sec:scan}

We perform random scanning of the model parameters
by imposing all the theoretical and experimental constraints.
The scanning ranges in the NS and IS are
\begin{align}
\label{eq:scan:range:1}
\hbox{NS: }& M_H \in \left[ 130, 2000 \right] \gev, \quad
M_A \in \left[ 15, 2000 \right] \gev,  \\ \nn
&  |\sba| \in \left[ 0.8, 1.0 \right], \quad
m_{12}^2\in \left[ 0, 1000^2 \right] \gev^2, 
\\[5pt] \nn
\hbox{IS: }& \mh \in \left[ 15,120 \right] \gev, \quad
M_A \in \left[ 15,2000 \right] \gev,  \\ \nn
&  |\sba| \in \left[ 0, 0.6 \right], \quad
m_{12}^2 \in \left[ 0, 1000^2 \right] \gev^2.
\end{align}
For the FCNC observables~\cite{Arbey:2017gmh,Misiak:2017bgg},
we take different ranges of $\mch$ and $\tb$ for type I/X and type II/Y:
\begin{align}
\label{eq:scan:range:2}
\hbox{type I/X: }&~
M_{H^\pm} \in \left[ 80, 2000 \right]  \gev, \quad
t_{\beta} \in \left[ 1, 50 \right],\\ \nn
\hbox{type II/Y: } &~
M_{H^\pm} \in \left[ 580, 2000 \right] \gev, \quad
t_{\beta} \in \left[ 0.5, 50 \right].
\end{align}

The range of $\sba$ is motivated by the current Higgs precision data~\cite{Aad:2019mbh}. 
And we scan over positive $m_{12}^2$
because we found in the preliminary scanning 
that a parameter point with negative $m_{12}^2$ does not 
satisfy the perturbativity, unitarity, or vacuum stability.
It is evident in $\lm_1$: see \eq{eq:quartic:NS:aligned}.
If $m_{12}^2<0$, the terms proportional to $\tb^2$ yield large $|\lm_1|$ and thus threaten the perturbativity,
especially for large $\tb$.
 In the preliminary check, 
we found that the vacuum stability condition is the most crucial factor in excluding the
parameter points with negative $m_{12}^2$.
If $m_{12}^2>0$, however,  the contribution from $M^2$ cancels that from $\mhh^2$.
For more efficient scanning, therefore,
only the positive values of $m_{12}^2$ are considered.

We randomly generate the six-dimensional parameter points in Eqs.~(\ref{eq:scan:range:1})
and (\ref{eq:scan:range:2}),
which are uniformly distributed.
Over the generated parameter points, we cumulatively impose the following steps:
\begin{description}
\item[Step-(i) Theory+FCNC:]
We require a parameter point to satisfy the theoretical stabilities and the FCNC results
using the public code \textsc{2HDMC}-v1.8.0~\cite{Eriksson:2009ws}.
		\ben
		\item Higgs potential being bounded from below~\cite{Ivanov:2006yq};
		\item Perturbative unitarity of the amplitudes of scalar-scalar, scalar-vector, and vector-vector scatterings at high energies~\cite{Kanemura:1993hm,Akeroyd:2000wc};
		\item Perturbativity of the quartic couplings~\cite{Branco:2011iw,Chang:2015goa};
		\item Vacuum stability~\cite{Barroso:2013awa};
		\item FCNC observables~\cite{Arbey:2017gmh,Misiak:2017bgg,HFLAV:2019otj}.
		
		\een
\item[Step-(ii) EWPD:]
      We calculate the Peskin-Takeuchi oblique parameters of $S$ and $T$ in the 2HDM~\cite{He:2001tp,Grimus:2008nb,Zyla:2020zbs},
      and compare them with the PDG and CDF results in Eqs.~(\ref{eq:STU:PDG})
      and (\ref{eq:STU:CDF}).
For two-parameter fitting
under the assumption of $U=0$, we require $\chi^2 < 5.99$.
\item[Step-(iii) RGEs for $\lmc>1\tev$:] 
We demand that the cutoff scale should be larger than 1 TeV.
Using the RGE in the 2HDM~\cite{Cheng:1973nv,Komatsu:1981xh,Branco:2011iw,Das:2015mwa,Basler:2017nzu}, we run the gauge couplings, the quartic couplings in the scalar potential, and the Yukawa couplings of 
the top quark, bottom quark, and tau lepton.
The initial conditions of the gauge couplings and the Yukawa couplings are
set at the top quark mass scale $m_t = 173.34\gev$~\cite{Oredsson:2018vio}.
We check the perturbativity, unitarity, and vacuum stability as increasing the energy scale.
If any condition is broken at the energy scale below 1 TeV,
we discard the parameter point.
We use the public code \textsc{2HDME}-v1.2~\cite{Oredsson:2018vio} at one-loop level.

\item[Step-(iv) Collider:] The collider constraints consist of two categories, 
the Higgs precision data and the direct search bounds at the LEP, Tevatron, and LHC.
To check the consistency with the Higgs precision data, 
we use \textsc{HiggsSignals}-v2.6.2~\cite{Bechtle:2020uwn}, which yields
  the $\chi^2$ output for 111 Higgs observables~\cite{Aaboud:2018gay,Aaboud:2018jqu,Aaboud:2018pen,Aad:2020mkp,Sirunyan:2018mvw,Sirunyan:2018hbu,CMS:2019chr,CMS:2019kqw}.
Since there are six model parameters,
the number of degrees of freedom is 105.
We demand that the $p$-value be larger than 0.05.
For consistency check with the direct searches at high energy colliders,
we use the public code \textsc{HiggsBounds}-v5.10.2~\cite{Bechtle:2020pkv}.
For each process,
we calculate the cross section in the model.  
When the model prediction is larger than the observed upper bound at 95\% C.L., we rule out the parameter point.
\end{description}

\begin{figure}[h] \centering
\begin{center}
\includegraphics[width=0.9\textwidth]{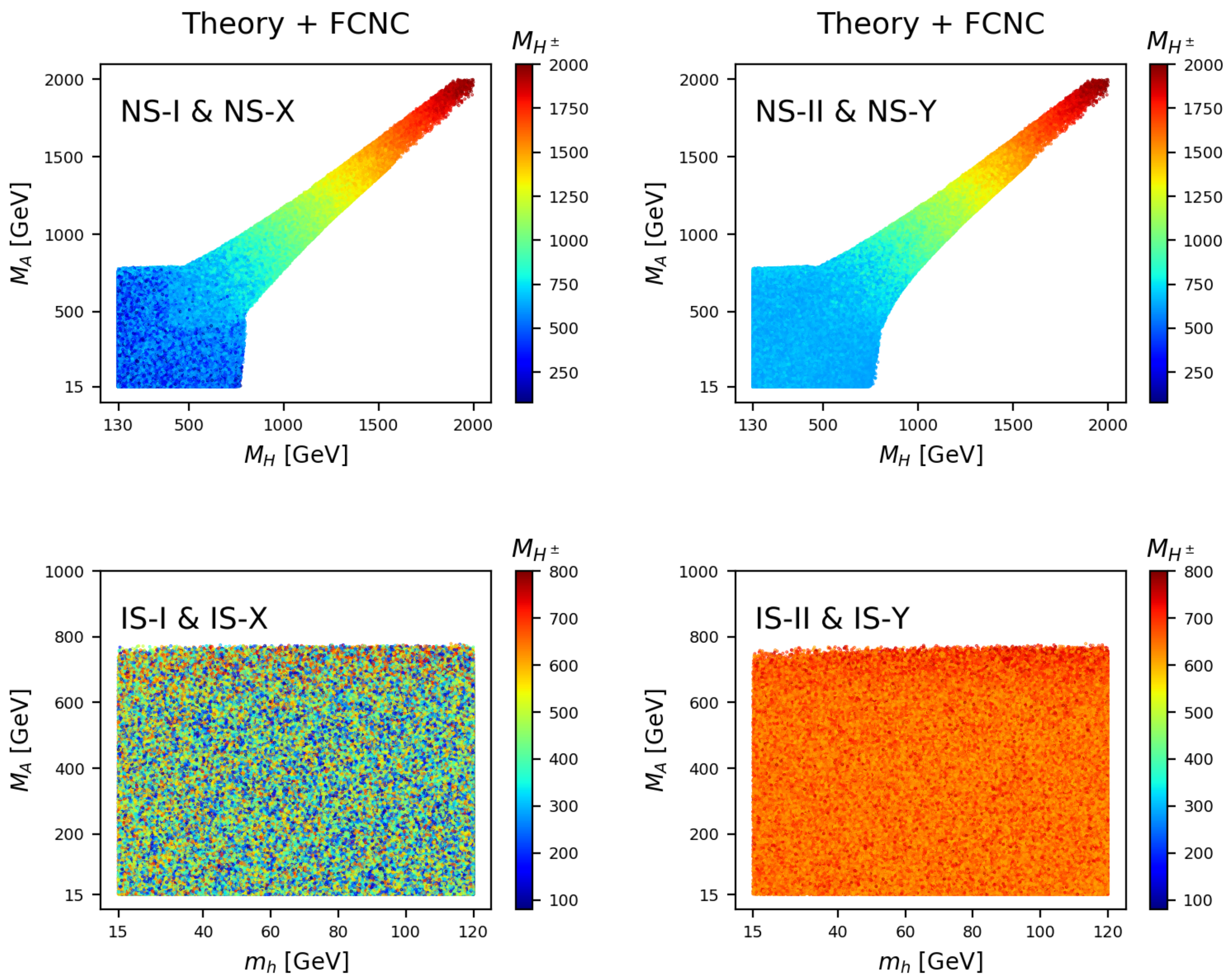}
\end{center}
\caption{\label{fig:step1}
Allowed regions of $(\mhh,\ma)$ in the NS (upper panels)
and $(\mh,\ma)$ in the IS   (lower panels) by the theoretical requirements and the FCNC observables.
The results in type I and type X are in the left panels, 
while those in type II and type Y are in the right panels.
The color codes indicate $\mch$.
}
\end{figure}

For each type in the NS and IS, we obtained $10^7$ parameter points that satisfy Step-(i),
which required to generate more than $10^{10}$ parameter points.
Before proceeding to the subsequent steps,
let us investigate the implications of Step-(i).
In Fig.~\ref{fig:step1},
we present $\ma$ versus $\mhh$ after Step-(i),  
where the color codes denote $\mch$.
The results in the NS (IS) are in the upper (lower) panels, and those at type I/X (type II/Y) are in the left (right) panels.
Figure \ref{fig:step1} obviously illustrates that
the theoretical requirements and the FCNC observables significantly restrict the masses of new Higgs bosons.
In the NS, Step-(i) demands very similar masses of new Higgs bosons
in the high mass region of $M_{A,H,\ch} \gsim 750\gev$,
while the low mass regions are uniformly permitted without correlations among the masses.
In the IS,
Step-(i) already puts the upper bounds on $\ma$ and $\mch$,
below about $750\gev$.

\begin{table}[!t]
\setlength\tabcolsep{10pt}
\centering
{\footnotesize\renewcommand{\arraystretch}{1.1} 
\begin{tabular}{|c|c||c|c|c|c||c|c|c|c|}
\toprule
 \multicolumn{2}{|c||}{} &\!\!\!\!\!\,Theory\,\!\!\!\!\!&  EWPD  & RGE & Collider & \!\!\!\!\!\,Theory\,\!\!\!\!\!&EWPD & RGE & Collider  \\ 
 \hline
 \multicolumn{2}{|c||}{type} & \multicolumn{4}{c||}{Normal scenario} & \multicolumn{4}{c|}{Inverted scenario} \\ \hline
 \multirow{2}{*}{I} 
 & PDG & $10^7$ & $1.3\times 10^6$ & $5.1\times 10^5$ & $6.0\times 10^4$ &$10^7$ & $7.2\times 10^5$ & $5.1\times 10^5$ & $8.5\times 10^4$ \\
  & CDF & $10^7$ & $4.4\times 10^5$  & $1.3\times 10^5$ & $1.4\times 10^4$  &$10^7$ & $1.3\times 10^5$ & $7.2\times 10^4$ & $1.9\times 10^4$\\
   \hline
 \multirow{2}{*}{II} 
 & PDG & $10^7$ & $1.1\times 10^6$ & $4.3\times 10^4$ & $2.0\times 10^4$ & $10^7$ & $2.1\times 10^5$ & 0 & 0\\
  & CDF & $10^7$ & $3.4\times 10^5$ & $3.0\times 10^3$ & $1.0\times 10^3$ &$10^7$ & $6.9\times 10^4$ & 0 & 0 \\
   \hline
\multirow{2}{*}{X}
& PDG & $10^7$ & $1.3\times 10^6$ & $5.1\times 10^5$ & $1.8\times 10^4$ &$10^7$ & $7.2\times 10^5$ & $5.1\times 10^5$ & $3.0\times 10^3$\\
  & CDF & $10^7$ & $4.4\times 10^5$ & $1.3\times 10^5$ & $3.0\times 10^3$ &$10^7$ & $1.3\times 10^5$ & $7.2\times 10^4$ & $1.0\times 10^3$\\
   \hline
 \multirow{2}{*}{Y} 
 & PDG & $10^7$ & $1.1\times 10^6$ & $4.3\times 10^4$ & $2.0\times 10^4$ &$10^7$ & $2.1\times 10^5$ & 0 & 0\\
  & CDF & $10^7$ & $3.4\times 10^5$ & $3.0\times 10^3$ & $1.0\times 10^3$ &$10^7$ & $6.9\times 10^4$ & 0 & 0\\
\bottomrule
\end{tabular}
}
\caption{The numbers of the parameter points that survive each step in the NS and IS for all the four types.
We linearly scan the parameters in Eqs.~(\ref{eq:scan:range:1}) and (\ref{eq:scan:range:2}).
For the EWPD,
we adopt two different schemes of the oblique parameters, 
without and with the CDF-updated $m_W$ measurement, denoted by \enquote{PDG} and \enquote{CDF}.
  }
\label{tab:survival}
\end{table}

Now we cumulatively impose the constraints of Step-(ii), Step-(iii), and Step-(iv).
In Table \ref{tab:survival},
we present the number of the parameter points that pass each step 
in the sixteen cases, the four types in the NS and IS without and with the CDF $m_W$ measurement.
We are well aware that just because a model has more surviving parameter points does not mean it is superior: 
Nature takes only one parameter point.
Nonetheless, the study of the surviving percentages
is meaningful in the situation where the experimentalists make every effort to find a new signal
without any information. 
A model with more allowed parameter points leaves more room for experimental exploration.
In addition, it is very important to present which constraint excludes which model more severely.  
The bottom-line is that the results in Table \ref{tab:survival} provide
the immediate comparison of 16 cases, 
coming to the main conclusions:
type I has the most surviving parameter points;
type II and type Y in the IS are excluded.

Brief comments on the dependence of the survival percentages on the scanning procedure are in order here.
We took the uniformly distributed samples over $\mch$, $\ma$, $M_{H/h}$, 
$\sba$, $\tb$, and $m_{12}^2$.
If the sampling were different, 
the results in Table \ref{tab:survival} would be different.
To estimate the changes,
we randomly scan over $m_{12}$, instead of $m_{12}^2$.
Scanning over $\log (m)$ is inappropriate 
since the upper bounds on the masses are only about $1\tev$.
The changes are below 10\% for type I/X and of the order of 10\% for type II/Y.
Consequently, the numbers in Table \ref{tab:survival}  are not physical, 
yet provide fair comparisons among all 16 cases.
However, the changes are not big enough to overturn the main conclusions in the comparative study of 16 cases,
such that type I has more parameter points than type X.
So, we discuss the physical implications of the CDF $m_W$ measurement,
based on the sampling over \eq{eq:scan:range:1} and \eq{eq:scan:range:2}.

Let us compare the overall differences between the NS and IS.
In the NS, 
all the types pass the final Step-(iv).
We find that type I has the most parameter points survived, while type II and type Y have the fewest.
In the IS, type II and type Y are excluded both in the PDG and CDF cases.\footnote{We confirmed
this conclusion by increasing the number of points in random scanning.}
Decisive is the combination of $\lmc>1\tev$ and the FCNC observables.
Since the former demands similar mass scales of the BSM Higgs bosons,
the light mass of $\mh$, below $125\gev$ by definition, necessitates light $\mch$.
Then the condition of $\mch>580\gev$ from $b\to s \gm$ prohibits the model. 
Type I and type X, on the other hand,
can accommodate the light charged Higgs boson without conflicting the FCNC observables.
So they are still allowed in the IS.

\begin{figure}[h] \centering
\begin{center}
\includegraphics[width=\textwidth]{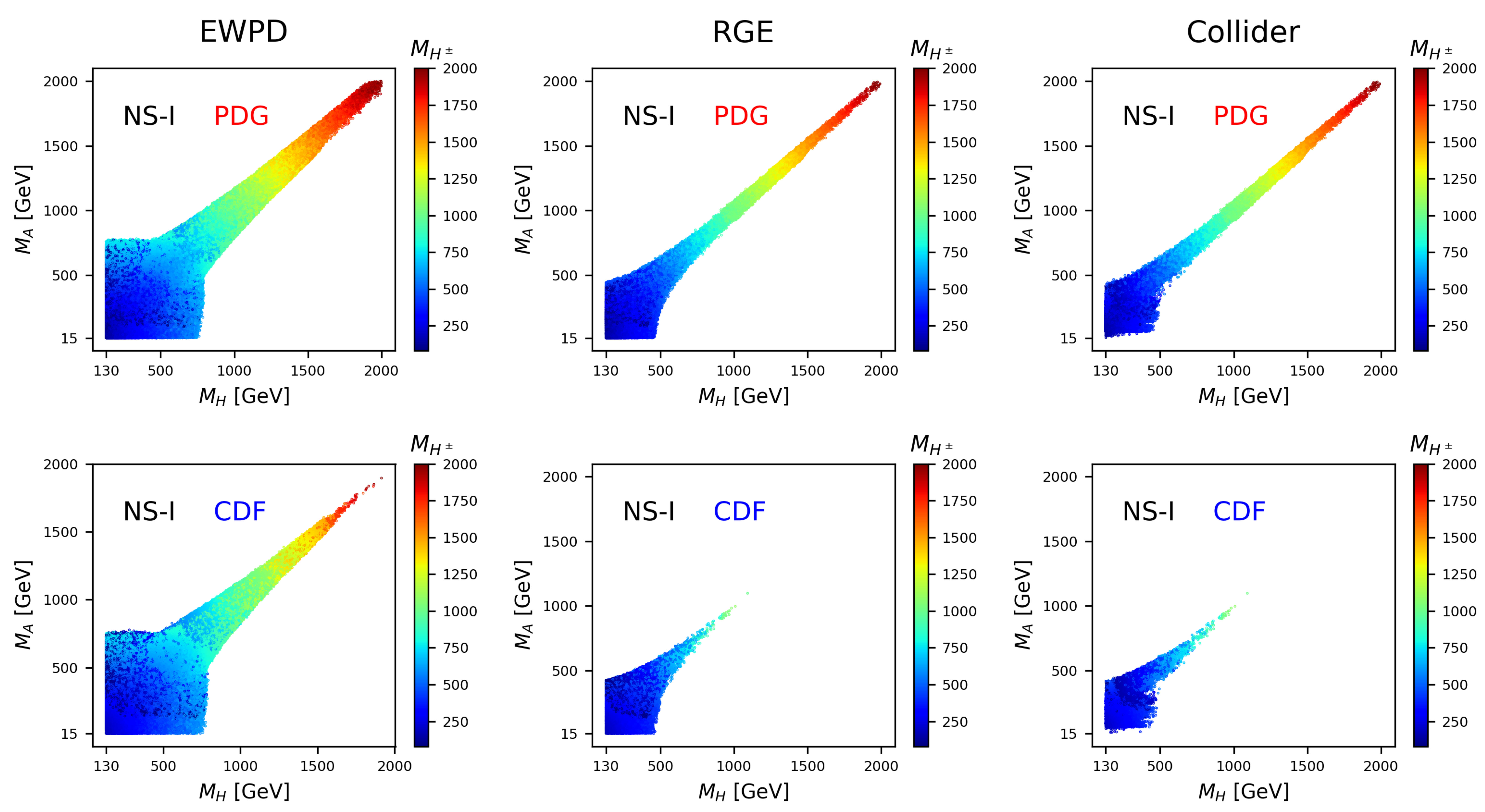}
\end{center}
\caption{\label{fig:PDG-CDF-MH-MA-NS-I}
In type I of the normal scenario,
the allowed parameter space of $(\mhh,\ma)$
after Step-(ii) (EWPD), Step-(iii) (Step(ii)+RGE), and Step-(iv) (Step(iii)+Higgs precision+Direct searches),
with the color code indicating $\mch$.
The upper panels are for PDG and the lower panels for CDF.
}
\end{figure}

Now we discuss the differences between the PDG and CDF cases.
The CDF $m_W$ allows fewer parameter points regardless of the type or the 
Higgs scenario:
the survival percentages are 
of the order of 0.01\% for the CDF, but of the order of 0.1\% for the PDG result.
The difference becomes evident from Step-(ii).
It is due to the tension that the EWPD needs sizable mass gaps in the CDF case
but $\lmc>1\tev$ favors the mass degeneracy~\cite{Das:2015mwa}.
To demonstrate this feature in more detail,
we present $\ma$ versus $\mhh$ at each step for the NS-I
in Fig.~\ref{fig:PDG-CDF-MH-MA-NS-I}.
The upper (lower) panels present the results for the PDG (CDF) case,
and the results after Step-(ii), Step-(iii), and Step-(iv) are
in the left, middle, and right panels, respectively.
At Step-(ii), the PDG and CDF cases yield similar funnel shapes in $(\mhh,\ma)$, stretching to the heavy mass regions.
But the size is different:
the CDF case permits a slimmer area with substantial mass gaps.
When imposing $\lmc > 1\tev$ at Step-(iii),
the difference between the PDG and CDF is stark.
The heavy mass region is excluded in the CDF case as discussed in Fig.~\ref{fig:ST:lm3:dMH:dMA}.
In summary, 
the combination of the $S/T$ constraint with $\lmc>1\tev$ 
puts the upper bounds on the masses of new Higgs bosons
in the CDF case.
Step-(iv) including the Higgs precision data and the direct search bounds
also reduces the parameter space,
both in the PDG and CDF cases.

\begin{figure}[h] \centering
\begin{center}
\includegraphics[width=\textwidth]{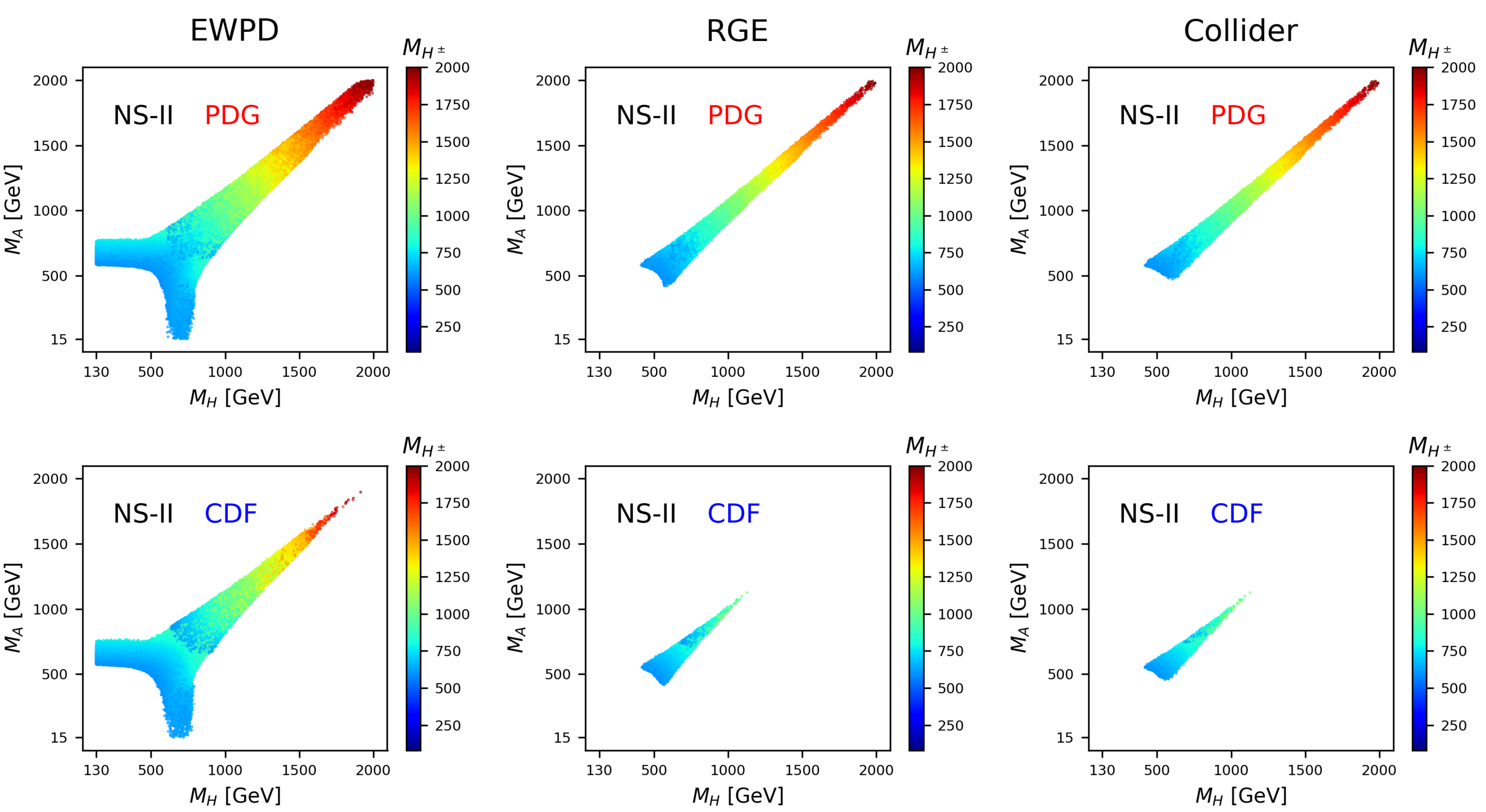}
\end{center}
\caption{\label{fig:PDG-CDF-MH-MA-NS-II}
In type II of the normal scenario,
the allowed parameter space of $(\mhh,\ma)$
after Step-(ii), Step-(iii), and Step-(iv),
with the color code indicating $\mch$.
The upper panels are for PDG while the lower panels for CDF.
}
\end{figure}

Figure \ref{fig:PDG-CDF-MH-MA-NS-II} presents the same results of $\ma$ versus $\mhh$ for type II.
As in type I, 
the EWPD yield similar shapes in $(\mhh,\ma)$ for the PDG and CDF cases,
and $\lmc> 1\tev$ at Step-(iii) puts the upper bounds on the masses of new Higgs bosons 
only in the CDF case.
Contrary to type I,
$\lmc> 1\tev$ also places the lower bounds on $\mhh$ and $\ma$ above about $580\gev$ 
both in the PDG and CDF cases.
It is attributed to the combination of $\mch\gsim 580\gev$ by $b\to s\gm$
with the similar mass scales of the BSM Higgs bosons by $\lmc>1\tev$.
At the final Step-(iv),
many parameter points are further excluded, 
but the mass bounds of new Higgs bosons remain almost intact.

The upper bounds on the new Higgs boson masses in the CDF case
have profound implications in the searches at the HL-LHC.
So, we present the allowed mass ranges in the CDF case for the NS
\begin{align}
 \begin{aligned}
 \label{eq:NS:parameters:CDF}
&\hbox{NS-I: } & &\mch\in[87, 1091]\gev,
\\  && &\mhh  \in [130,1092]\gev, &  &\ma \in [22,1098]\gev;
\\[5pt]
&\hbox{NS-II: }  & &\mch\in[598,1139]\gev,
\\  && &\mhh \in [419,1128]\gev, &  & \ma \in [459,1125]\gev;
\\[5pt]
&\hbox{NS-X: } & &\mch\in[99,1091]\gev,
\\ && & \mhh  \in [132,1092]\gev, &  &\ma \in [30, 1098]\gev;
\\[5pt]
& \hbox{NS-Y: }  & &\mch\in[595,1139]\gev,
\\  && & \mhh  \in [419,1128]\gev, &  &\ma \in [459,1125]\gev;
\end{aligned}
 \end{align}
 and for the IS
 \begin{align}
 \begin{aligned}
 \label{eq:IS:parameters:CDF}
&\hbox{IS-I: }  & &\mch\in[144,455]\gev,
\\  && & \mhh \in [16,120]\gev, &  &\ma \in [38,429]\gev;
\\[5pt] 
& \hbox{IS-X: } & &\mch \in[166,446]\gev,
\\  && &\mhh  \in [62.5,120]\gev, &  &\ma \in [16,420]\gev.
\end{aligned}
 \end{align}

Finally, we want to discuss the possibility of light BSM Higgs bosons in the CDF case.
First, a light charged Higgs boson with a mass below the top quark mass is feasible in type I and type X.
It is consistent with
the current searches for the light charged Higgs boson at the LHC via the production of $t\to \ch b$
in the decay modes of $\ch\to \tau^\pm\nu$~\cite{ATLAS:2018gfm,Sirunyan:2019hkq},
$\ch \to cb$~\cite{ATLAS:2021zyv,CMS:2018dzl},
and $\ch \to cs$~\cite{ATLAS:2013uxj,CMS:2015yvc,CMS:2020osd}.
For future searches,
the detailed characteristics of the light $\ch$ is important.
In the IS,
the mass difference of the light $\mch$ from the top quark mass is small:
see \eq{eq:IS:parameters:CDF}.
The soft $b$ jet in the process of $t\to \ch b$ makes it challenging
to observe the light $\ch$ through the conventional production channel~\cite{CMS:2013xio,Craig:2016ygr}.
Bosonic productions of the light charged Higgs boson deserve to pursue~\cite{Cheung:2022ndq}.
In the NS, type I and type X accommodate a larger mass gap between $m_t$ and $\mch$
but the $b$ jet from $t\to \ch b$ is still too soft to enjoy high $b$-tagging efficiency.

Second, a light pseudoscalar at a mass below $62.5\gev$ is allowed in type I and type X.
Although it seems to contradict the current data on the exotic Higgs decay of $ \hsm\to AA$,
the final parameter points pass the \textsc{HiggsSignals} and \textsc{HiggsBounds},
especially
the CMS searches for $\hsm\to AA$ in the final states of $2\mu 2\tau /4\tau$~\cite{CMS:2019spf},
$4\tau/ 2\mu 2b/ 2\mu2\tau$~\cite{CMS:2017dmg},
and $2\mu2\tau $~\cite{CMS:2018qvj},
as well as 
the LEP search for $\ee\to 4b Z/4\tau Z$~\cite{ALEPH:2006tnd}.
The main reason is small trilinear coupling $\hat{\lm}_{hAA}$, 
which is defined by $\lg \supset (1/2) v  \hat{\lm}_{hAA} \hsm AA$.
In the NS, $\hat{\lm}_{hAA}$ is~\cite{Kim:2022hvh}
\begin{align}
\hat{\lm}_{hAA}^{\rm NS} &= \lf 2 M^2 - 2\ma^2-\mh^2 \ri \sba +(\mh^2 -M^2)\lf \tb -\frac{1}{\tb} \ri \cba.
\end{align}
If $M^2 \simeq \ma^2 + \mh^2/2$,
$\hat{\lm}_{hAA}$ and thus $\br(\hsm\to AA)$ are suppressed.
We found that the finally allowed parameters
accommodate $\br(\hsm\to AA)\lsim 0.22$ in type I and $\br(\hsm\to AA)\lsim 0.11$ in type X.
Larger $\br(\hsm\to AA)$  in type I is explained by
the dominant decay of $A\to bb$ with the branching ratio of about 80\%,
which invalidates the main search modes of $A\to 2\tau/2\mu$.

\section{Characteristic features of the normal scenario}
\label{sec:NS}

In this section,
we study the characteristics of the finally allowed parameter points in the NS.
Since the results of type I (type II) are similar to those of type X (type Y),
we show type I and type X (type II and type Y) together.

\subsection{Type I and type X}

\begin{figure}[h] \centering
\begin{center}
\includegraphics[width=0.85\textwidth]{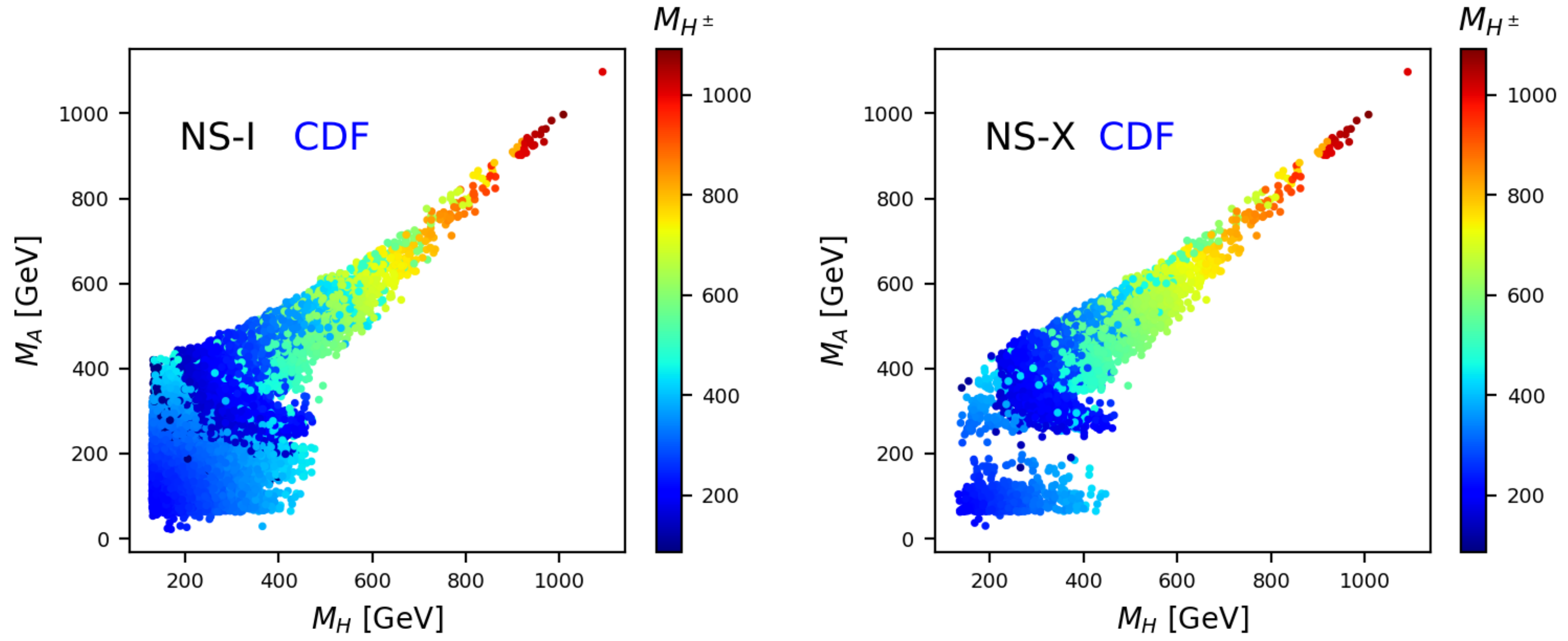}
\end{center}
\caption{\label{NS-IX-MH-MA-McH.pdf}
$\ma$ versus $\mhh$ with a color code of $\mch$ in type I (left panel) and type X (right panel) in the CDF case.
We focus on the normal scenario.}
\end{figure}

To take a closer look at the allowed masses of the BSM Higgs bosons at the final Step-(iv),
we present  $\ma$ versus $\mhh$ with a color code of $\mch$ for type I (left panel) and type X (right panel)
in Fig.~\ref{NS-IX-MH-MA-McH.pdf}.
In the heavy mass region with $M_{H,A} \gsim 600\gev$,
a correlation exists among the masses of new Higgs bosons.
Both $\dmhh$ and $\dma$ are clustered around $\Dt M_{H,A} \simeq -100\gev$.
For $M_{H,A} \lsim 600\gev$, the mass correlations are weak.

\begin{figure}[h] \centering
\begin{center}
\includegraphics[width=0.9\textwidth]{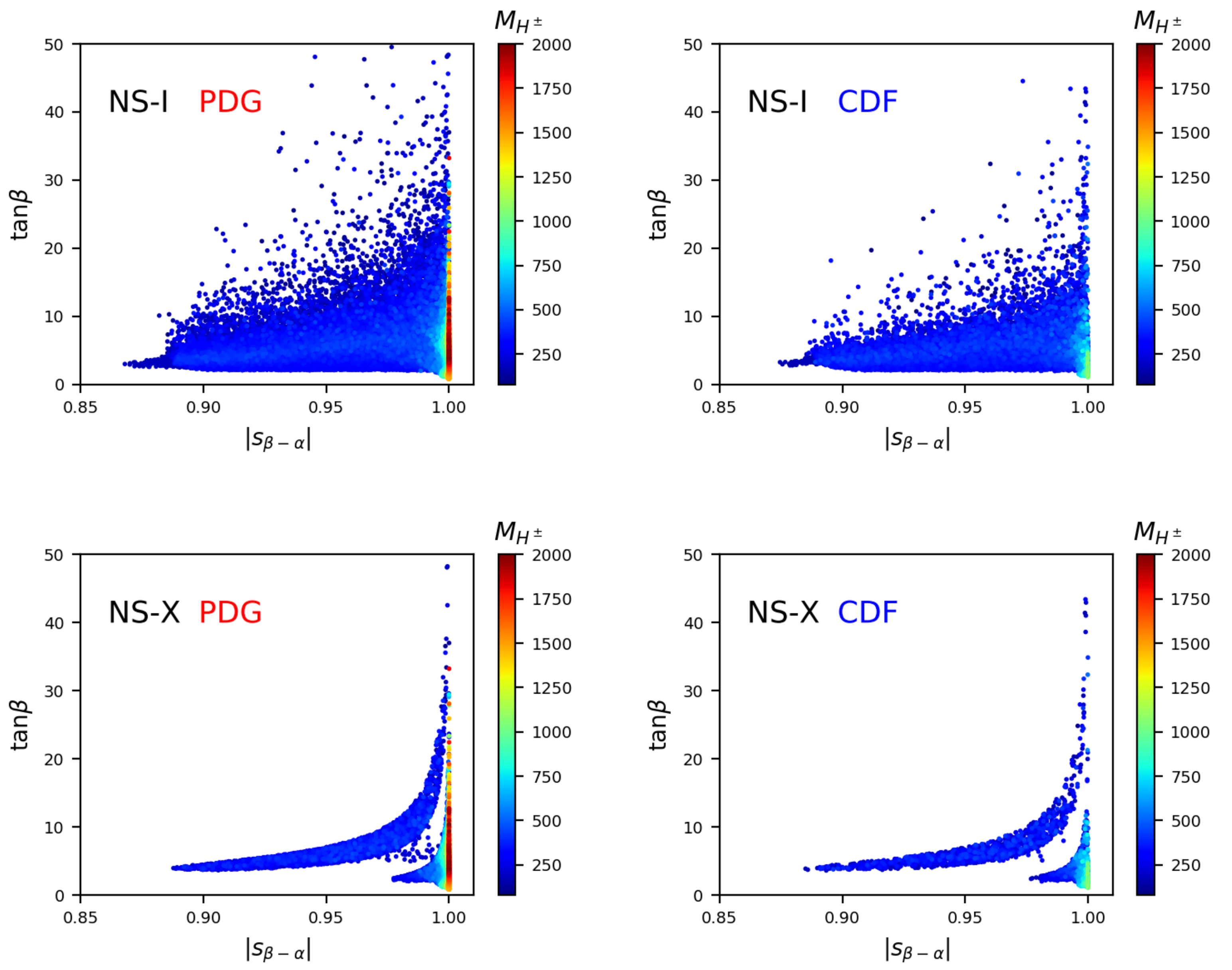}
\end{center}
\caption{\label{fig:sba-tb-McH-NS-IX}
Allowed regions of $( |\sba|, \tb)$ in type I (upper panels) and type X (lower panels),
with a color code indicating $\mch$ for the normal scenario.
We compare the results before (left panels) and after (right panels) the CDF $m_W$ measurement.
}
\end{figure}

Now we move on to the couplings.
Figure \ref{fig:sba-tb-McH-NS-IX} presents $\tb$ versus $|\sba|$ in type I (upper panels) and type X (lower panels),
with a color code indicating $\mch$.
We compare the results in the PDG case (left panels) with those in the CDF case (right panels).
The Higgs precision data play a vital role in limiting $|\sba| \gsim 0.88$,
which is similar in all the four panels.
But the distributions of $\tb$ are noticeably different according to the type and $m_W$.
In type I, $\tb$ is more spread out than in type X.
Inside type I, the PDG and CDF show dissimilar patterns:
$\tb$ in the CDF case is clustered in smaller value region than in the PDG case.
Since all the fermion Yukawa couplings to $H$, $A$, and $\ch$ in type I are inversely proportional to $\tb$
in the Higgs alignment limit,
small $\tb$ increases the LHC discovery potential of new Higgs bosons
through fermionic production and decay channels.
In type X, a correlation between $\tb$ and $|\sba|$ is strong.
There exists an upper bound on $\tb$ when the Higgs alignment is broken even a little:
for instance,  $\tb<7$ if $|\sba|=0.95$.

\begin{figure}[h] \centering
\begin{center}
\includegraphics[width=0.47\textwidth]{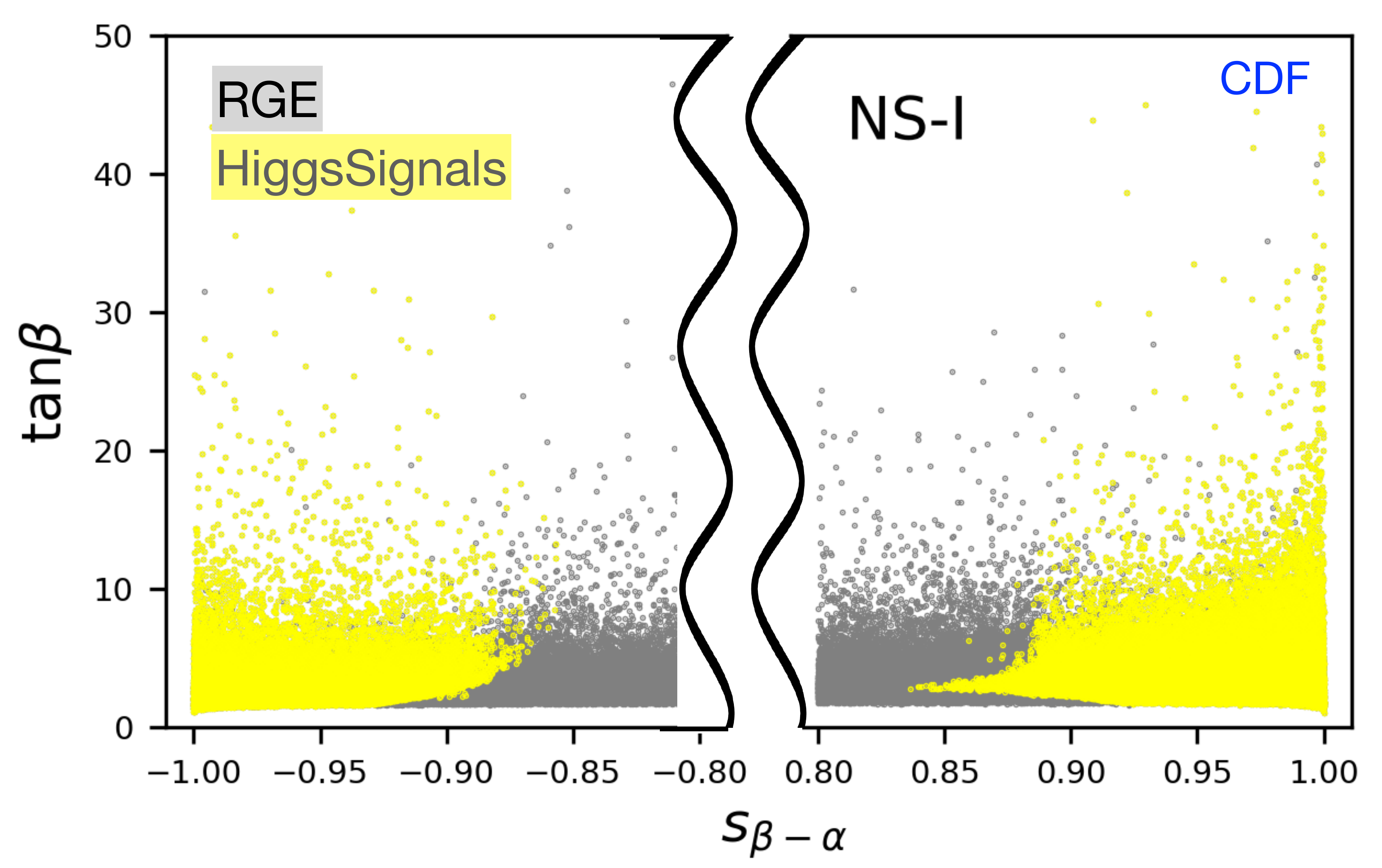}~~
\includegraphics[width=0.47\textwidth]{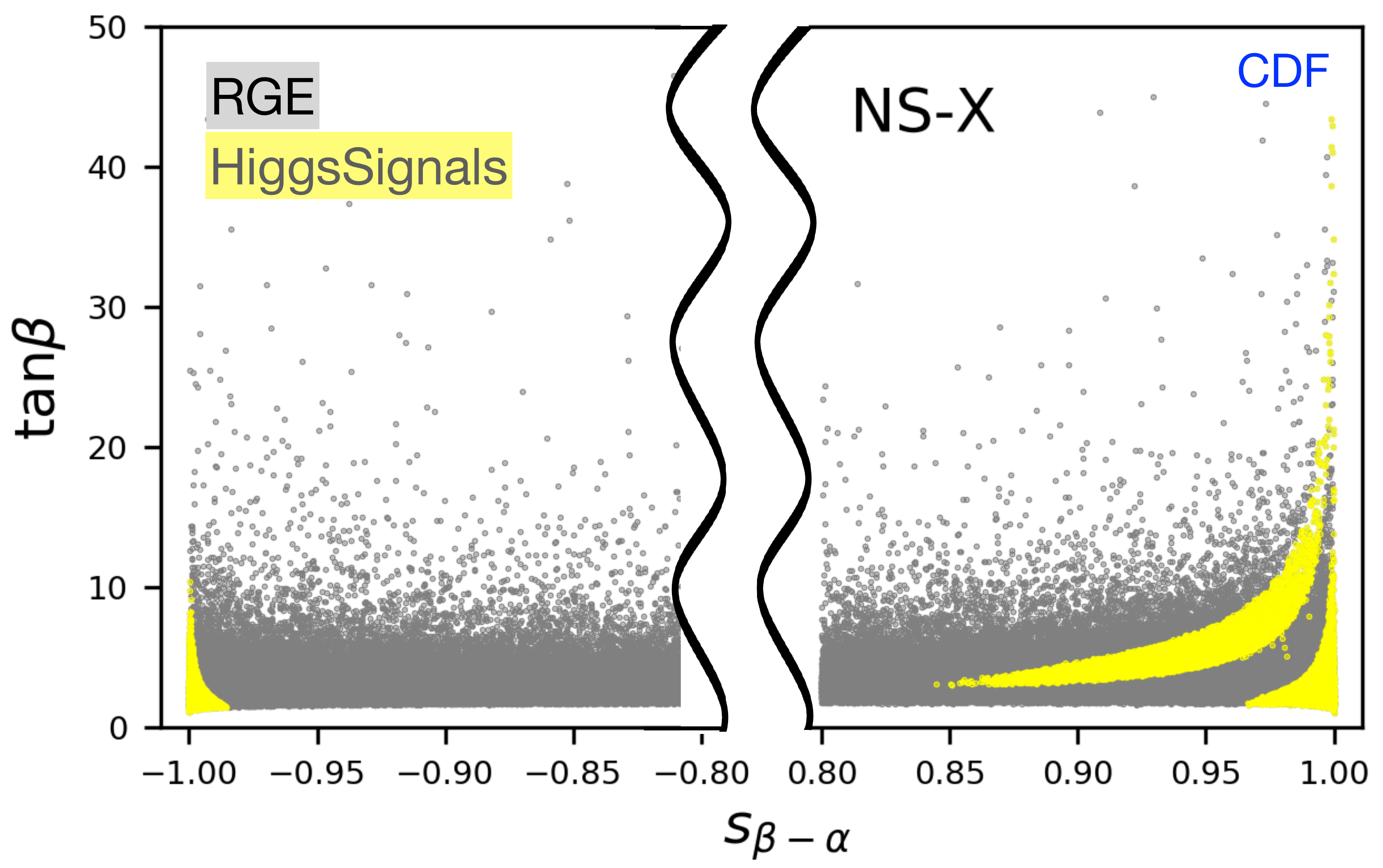}
\end{center}
\caption{\label{fig:NS-IX-tb-sba}
$\tb$ versus $ \sba$ in type I (left panel) and type X (right panel)
of the normal scenario
that pass the RGE (gray) and the Higgs precision data (yellow).
}
\end{figure}

Another observation in Fig.~\ref{fig:sba-tb-McH-NS-IX} is the \enquote{arm} region in type X.
It is due to the different effects of the Higgs precision data
on the positive and negative $\sba$ regions.\footnote{Note that negative $\sba$ in our scheme corresponds to negative $\cba$ in the positive $\sba$ scheme.}
In Fig.~\ref{fig:NS-IX-tb-sba},
we separately display, over $(|\sba|,\tb)$, the parameter points with $\lmc>1\tev$ (gray)
and those additionally satisfying the Higgs precision data (yellow):
yellow points are on top of gray ones.
The results of NS-I are in the left panel
and those of NS-X are in the right panel.
At Step (iii) with the RGE analysis,
the allowed region for $\sba$ is almost symmetric about $\sba=0$.
When imposing the Higgs precision data,
type I keeps the symmetric shape but type X does not.
In type X, most of the parameter space with negative $\sba$ is excluded, except for the Higgs alignment limit.

The presence or absence of the arm region is determined by the tau lepton Yukawa coupling to the Higgs boson $h$:
\begin{align}
\label{eq:tau:Yukawa:I}
\hbox{NS-I: } &~ \xi^h_\tau = \frac{\ca}{\sb} = \sba + \frac{\cba}{\tb},
\\ \label{eq:tau:Yukawa:X}
\hbox{NS-X: } &~ \xi^h_\tau = -\frac{\sa}{\cb} = \sba - \tb \cba.
\end{align}
In type I, the $\cba$ term is suppressed by large $\tb$ so $\sba \simeq -1$ does not change $ |\xi^h_\tau|$ much.
This is why negative $\sba$ satisfies the Higgs precision data in type I.
In type X, however,
the $\cba$ term is proportional to $\tb$.
If $\sba=-1 + \es$ and $\es\neq0$,
large enough $\tb$ overly increases $ |\xi^h_\tau|$ so that
the branching ratio of $h\to \ttau$ exceeds the experimental bound.
So, most of the negative $\sba$ region is removed.
The \enquote{arm} region in type X,
which appears for $\sba>0$,
is also explained by \eq{eq:tau:Yukawa:X}.
When $\tb \simeq  2/\cba$, $\xi^h_\tau $ approaches $-1$ for $\sba\simeq 1$.
We have the wrong-sign Yukawa coupling for the tau lepton.

\subsection{Type II and type Y}

\begin{figure}[h] \centering
\begin{center}
\includegraphics[width=1\textwidth]{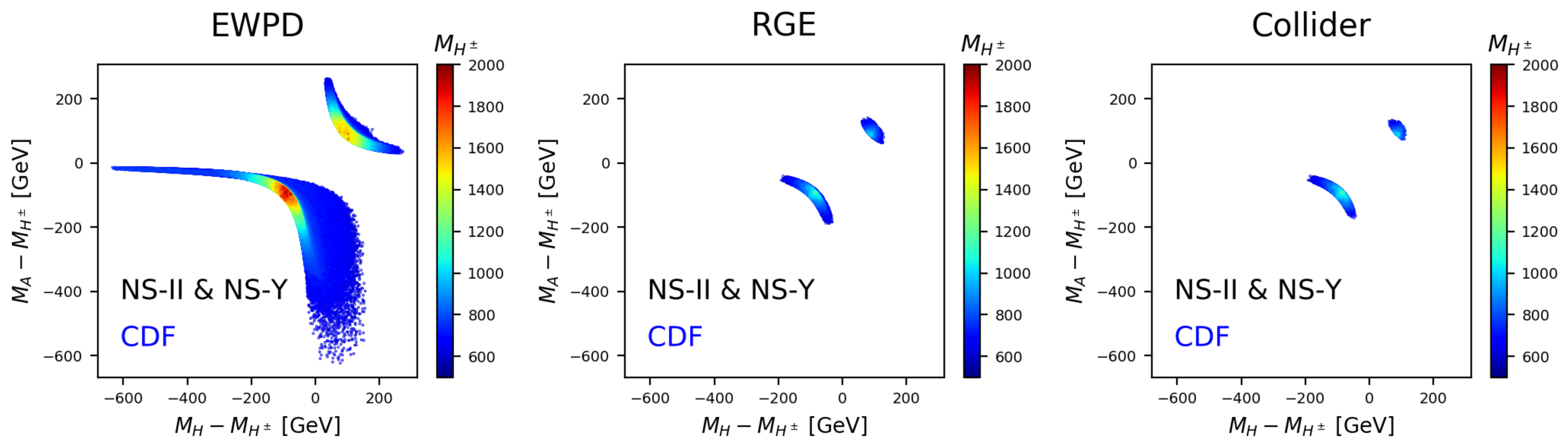}
\end{center}
\caption{\label{fig:NS-IIY-dMH-dMA-McH}
In type II and type Y of the normal scenario, 
the allowed parameter points of $( \dmhh, \dma)$ at the Step-(ii) in the left panel,
at the Step-(iii) in the middle panel, and at the Step-(iv) in the right panel, 
where $\Dt m \equiv m-\mch$.
The color code denotes $\mch$.
}
\end{figure}

We first point out that the allowed parameter points at each step in type II are almost the same as those in type Y.
So, all the results in this subsection are common for type II and type Y.
In type II and type Y with the CDF $m_W$, the biggest impact comes from 
the condition of $\lmc>1\tev$.
Figure \ref{fig:NS-IIY-dMH-dMA-McH} shows $\dma$ versus $\dmhh$ at Step-(ii) in the left panel,
at Step-(iii) in the middle panel, and at Step-(iv) in the right panel.
The left panel shows that the oblique parameters of $S_{\rm CDF}$ and $T_{\rm CDF}$ permit the hyperbola-shape
with a sufficiently large mass gaps.\footnote{The negative $\Dt M_{H,A}$ region
is different from Fig.~\ref{fig:ST:lm3:dMH:dMA},
because we assumed the Higgs alignment limit only in Fig.~\ref{fig:ST:lm3:dMH:dMA}.}
Imposing $\lmc>1\tev$ (middle panel) excludes a large portion of the parameter space,
particularly with $|\Dt M_{H,A}|\gsim200\gev$.
It is because too large mass gaps invoke fast running of the quartic couplings,
resulting in the failure of the unitarity and vacuum stability at the energy scale below $1\tev$.
The area near the mass degeneracy of $\mhh=\ma=\mch$ is also removed.
Lastly, the constraints from the collider data do not considerably change $ \dma$ versus $\dmhh$.

\begin{figure}[h] \centering
\begin{center}
\includegraphics[width=0.85\textwidth]{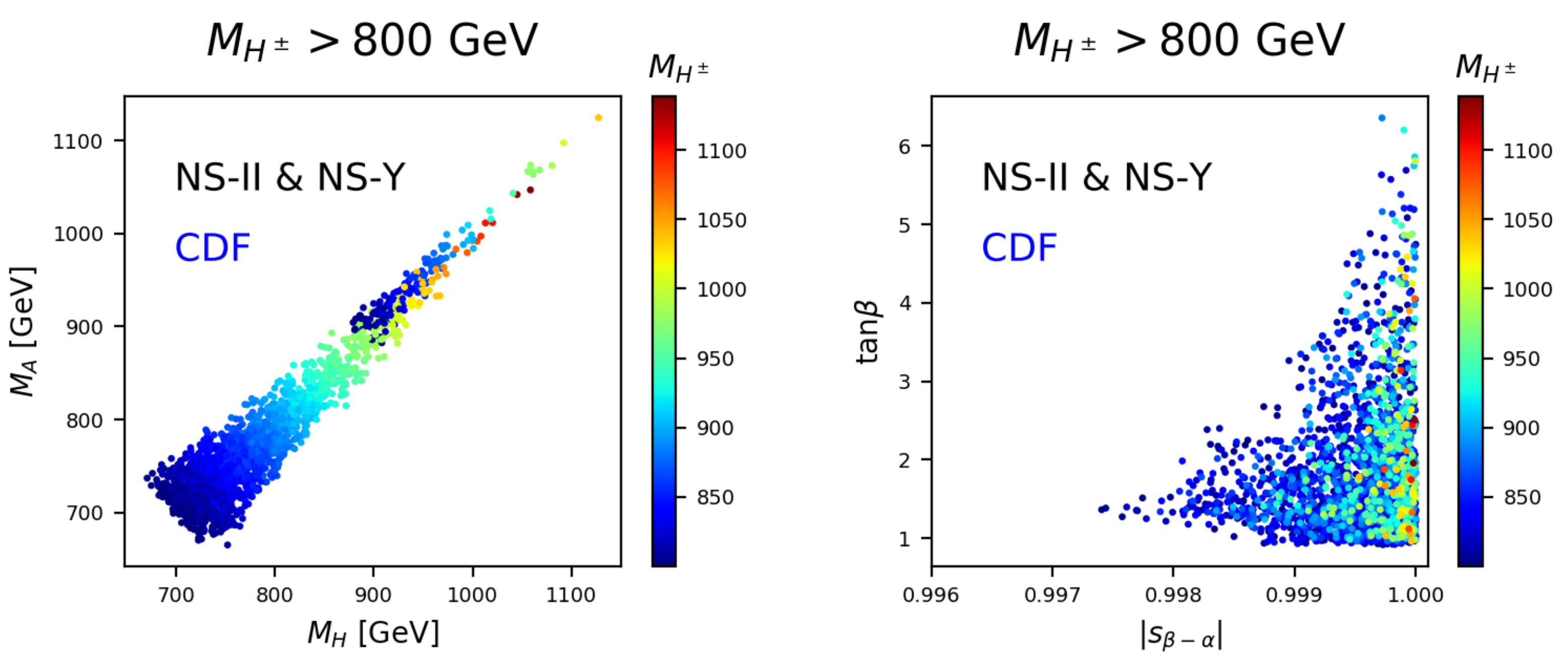}
\end{center}
\caption{\label{fig:NS-IIY-McH800}
For type II in the normal scenario with $\mch>800\gev$, 
the allowed parameter points of $( \mhh, \ma)$ in the left panel
and $(|\sba|,\tb)$ in the right panel.
The color code denotes $\mch$.
}
\end{figure}

Finally, let us discuss the constraint from $b\to s \gm$.
In the main analysis, we took a conservative bound on the charged Higgs boson mass as $\mch > 580\gev$
in type II and type Y.
But the bound considerably increases to about 800 GeV
if we adopt the calculation of the NNLO QCD corrections to $\br(\bar{B} \to X_s \gm)$ in the SM 
without the interpolation in the charm quark mass~\cite{Misiak:2020vlo}.
The stronger condition on $\mch$ restricts the other model parameters further.
Focusing on $\mch>800\gev$,
we additionally generated parameter points.
Figure~\ref{fig:NS-IIY-McH800} presents $\ma$ versus $\mhh$ in the left panel,
and $\tb$ versus $|\sba|$ in the right panel, for $\mch>800\gev$.
The color code indicates $\mch$.
The lower bounds on $\mhh$ and $\ma$ increase into about $670\gev$.
Although they are smaller than the lower bound on $\mch$,
the heavy masses of the BSM Higgs bosons make it challenging to probe NS-II or NS-Y at the HL-LHC.
The right panel in Fig.~\ref{fig:NS-IIY-McH800} exhibits that
the constraints on $|\sba|$ and $\tb$ are stronger for $\mch>800\gev$.
The Higgs alignment is almost exact
and the value of $\tb$ is intermediate like $ \in [0.9,\,6.4]$.

\section{Characteristic features of the inverted scenario}
\label{sec:IS}

\begin{figure}[h] \centering
\begin{center}
\includegraphics[width=0.98\textwidth]{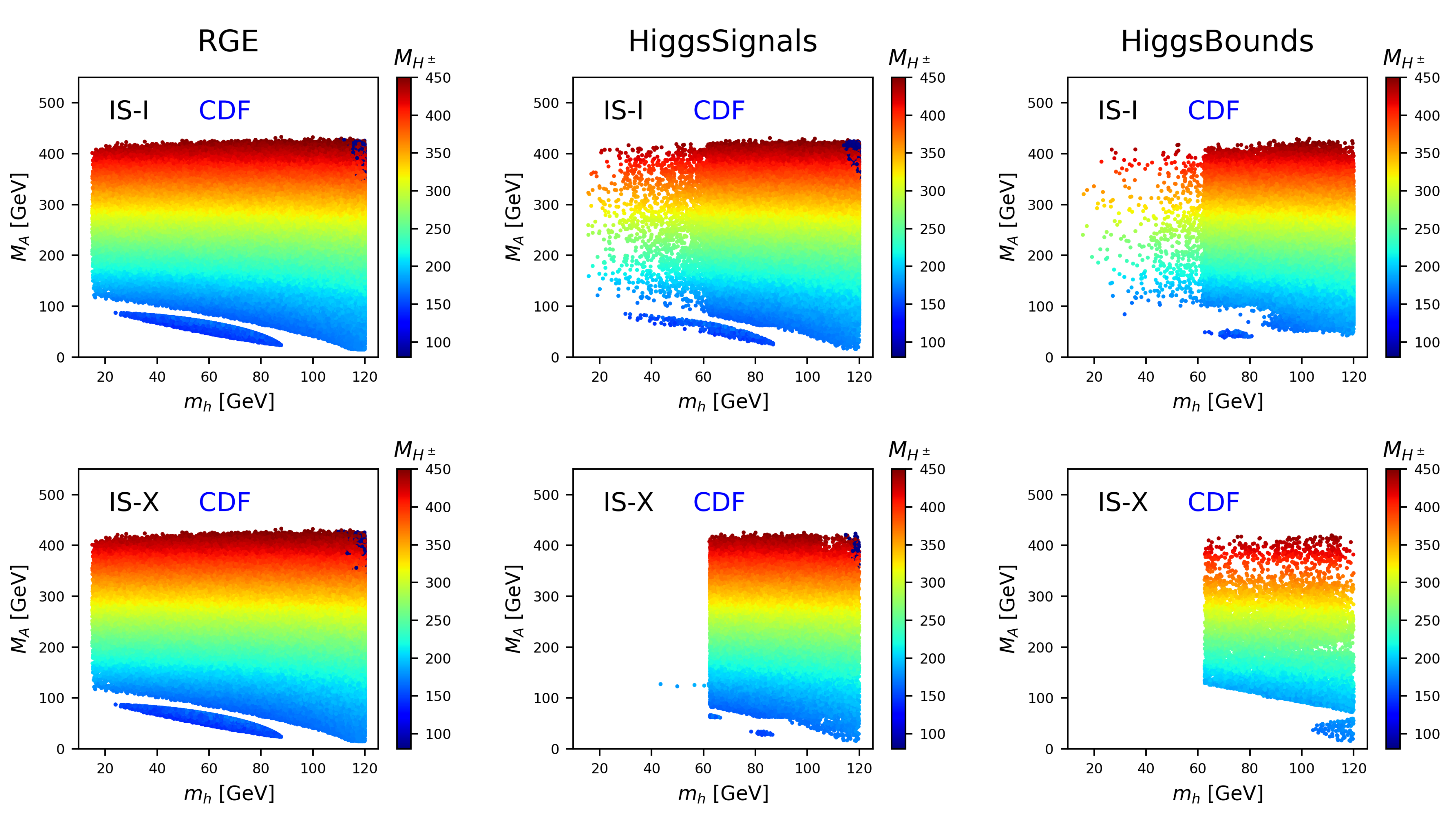}
\end{center}
\caption{\label{fig:IS-IX-mh-MA-McH}
Allowed parameter points of $( \mh, \ma)$ after imposing $\lmc>1\tev$ (left panels),
the Higgs precision data (middle panels), and the direct search bounds (right panels)
for type I (upper panels) and type X (lower panels) in the inverted scenario.
The color code denotes $\mch$.
}
\end{figure}

The IS accommodates a light Higgs boson at a mass below $125\gev$.
This exotic scenario has drawn a lot of interest since it satisfies the theoretical requirements
and the experimental data.
However, the RGE analysis changes this conclusion,
which has not yet been performed for the IS.
According to our RGE study, type II and type Y in the IS do not retain the stability of the scalar potential
up to $1\tev$,
which are excluded by the condition of $\lmc>1\tev$.
In this section, therefore, we investigate the characteristics of
the finally allowed parameter points of type I and type X in the CDF case.

The first remarkable feature is considerably different survival percentages between type I and type X:
see Table \ref{tab:survival}.
In type X, only about 0.01\% of the parameter points at Step-(i) are finally allowed,
but in type I, the number is 0.19\%.
To find the origin,
we show, in Fig.~\ref{fig:IS-IX-mh-MA-McH}, $\ma$ versus $\mh$
at the steps of $\lmc>1\tev$ (left panels),
the Higgs precision data (middle panels),
and the direct search bounds (right panels).
The results of type I  are in the upper panels and those of type X are in the lower panels.
At the RGE step, type I and type X yield almost the same results.
A significant difference arises after imposing the Higgs precision data.
In type X, most of the parameter points with $\mh\lsim 62.5\gev$
are excluded unlike in type I.
The direct search bounds further widen the difference between type I and type X.
The leptophilic nature of type X, being more severe for large $\tb$,
brings out the severe restriction.

The second noteworthy feature in Fig.~\ref{fig:IS-IX-mh-MA-McH} is that type I permits $\mh<\mhsm/2$,
but type X does not.
This is due to the different decay modes of the light $h$ in type I and type X.
In type I where the ratios of the Yukawa couplings of $h$ are the same as in the SM,
the light $h$ dominantly decays into a pair of $b$ quarks
with the branching ratio of about 80\%.
Since the searches for a light Higgs boson at the LHC make use of the $4\tau$ and $2\mu2\tau$ states~\cite{CMS:2017dmg,CMS:2018qvj,CMS:2019spf},
type I is less constrained.
On the contrary,
$h$ in type X decays dominantly into $\ttau$,
which is strongly limited from the $4\tau/2\mu2\tau$ final states.

\begin{figure}[h] \centering
\begin{center}
\includegraphics[width=0.5\textwidth]{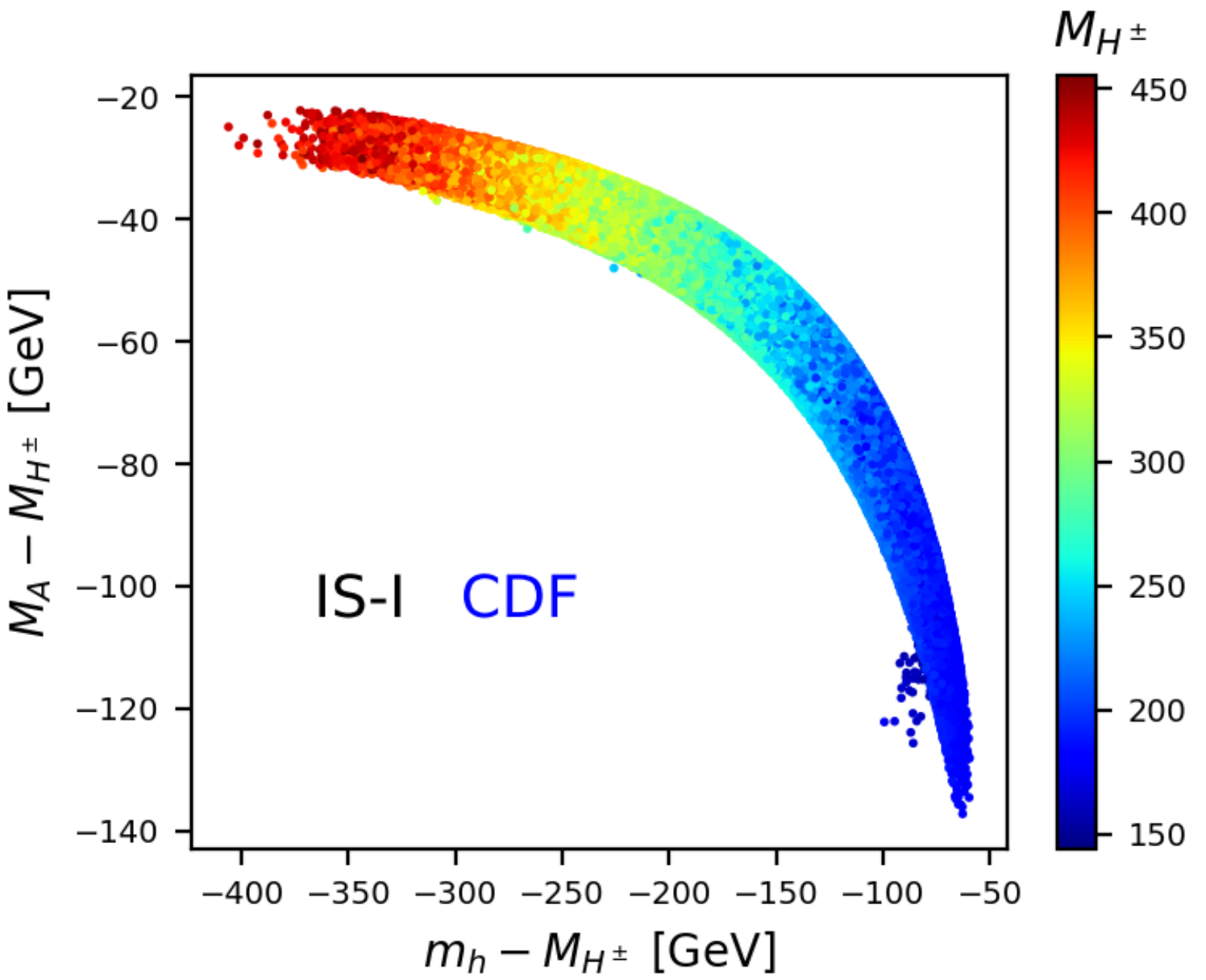}
\end{center}
\caption{\label{fig:IS-I-dmh-dMA-McH}
Allowed parameter points of $( \dmh, \dma)$ at the final step
for type I in the inverted scenario, with a color code of $\mch$.
}
\end{figure}

Another important result in the IS with the CDF $m_W$ is the strong correlation among $\ma$, $\mh$, and $\mch$.
Figure \ref{fig:IS-I-dmh-dMA-McH} shows $\dma$ versus $\dmh$ 
with a color code of $\mch$ for IS-I,
which is similar to IS-X.
We first notice that the IS in the CDF case allows only the negative $\dmh$ and the negative $\dma$.
It is to be compared with the NS in the CDF case,
which also permits $\dmh>0$ and $\dma>0$: see Fig.~\ref{fig:NS-IIY-dMH-dMA-McH}.
The sign of $\dma$ has a big impact on the bosonic decays of the BSM Higgs bosons~\cite{Arhrib:2017wmo,Mondal:2021bxa,Kanemura:2011kx,Arhrib:2021xmc,Arhrib:2021yqf,Cheung:2022ndq}.
Since the charged Higgs boson is heavier than the pseudoscalar,
$\ch\to A W^{\pm(*)} $ is feasible but $A\to \ch W^{\mp(*)}$ is not.
Another intriguing aspect is the approximate mass degeneracy of $\mch\simeq \ma$ for heavy $\mch$:
if $\mch \gsim 350\gev$, $| \dma|<40\gev$.
We expect that this region is very difficult to probe at the LHC.
Pair productions of $A H$, $A \ch$, and $H^+ H^-$ has kinematic suppression by heavy masses.
The gluon fusion production of $A$ or $h$ is suppressed by large $\tb$.
Furthermore, the bosonic decay mode of $\ch$ accompanies the soft fermions from the off-shell $W$.

\begin{figure}[h] \centering
\begin{center}
\includegraphics[width=0.85\textwidth]{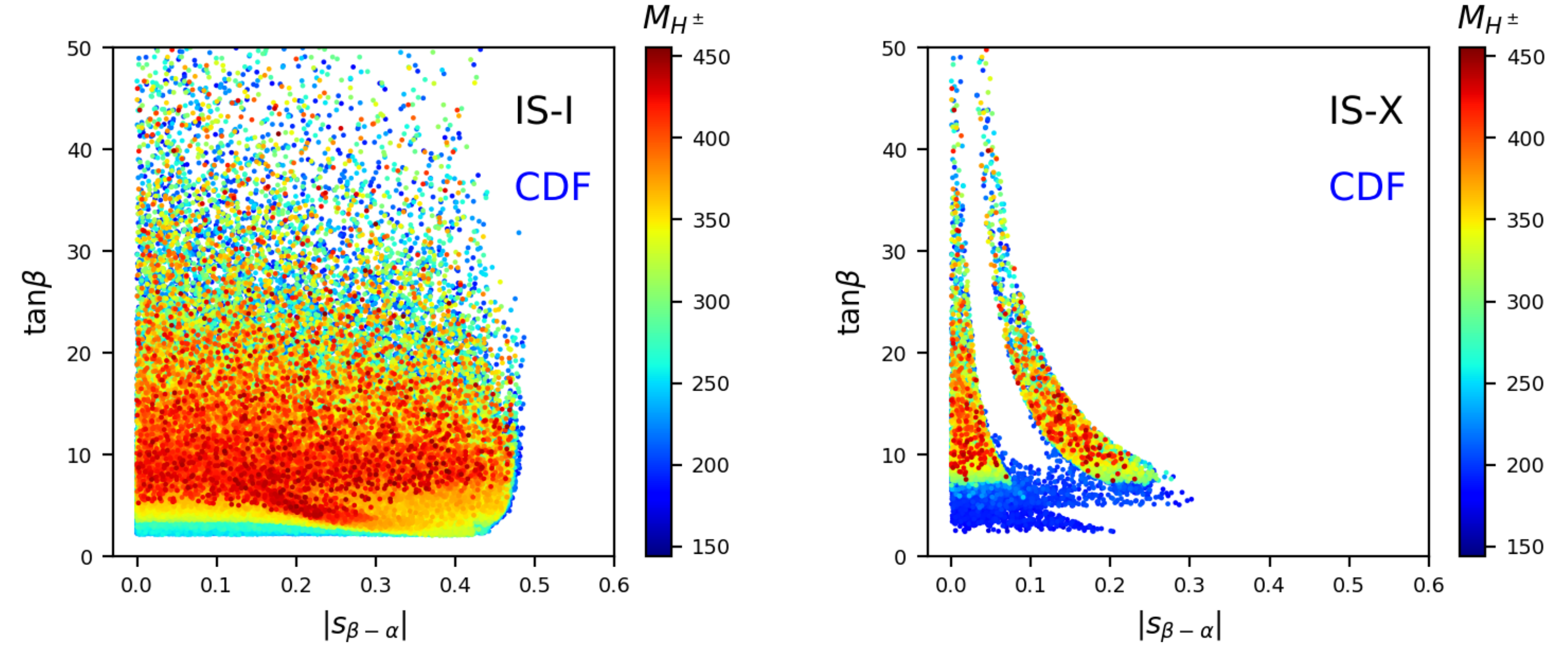}
\end{center}
\caption{\label{fig:IS-IX-sba-tb-McH}
Allowed parameter points of $( |\sba|, \tb)$ at the final step
for type I (left panel) and type X (right panel) in the inverted scenario, with a color code of $\mch$.
}
\end{figure}

Finally, we present $ \tb$ versus $ |\sba|$  in Fig.~\ref{fig:IS-IX-sba-tb-McH}
for IS-I (left panel) and IS-X (right panel) for the CDF case.
The difference between type I and type X is considerable.
In type I, a sizable deviation from the Higgs alignment limit is feasible.
The value of $|\sba|$ can be as large as about 0.48.
In addition, 
large $\tb$ around 50 is also permitted, irrespective of $\sba$.
In type X, $|\sba| \lsim 0.3$: the maximum deviation from the Higgs alignment is smaller than that in type I.
In addition, $|\sba|$ and $\tb$ are more strongly correlated:
if $|\sba| =0.3$,  $\tb$ is almost fixed to be 5.5.

\section{Conclusions}
\label{sec:conclusions}

The recent $W$-boson mass measurement by the CDF collaboration
has a significant impact on new physics models.  
The central values of the Peskin-Takeuchi parameters of $S$ and $T$  
shift to larger values: $S = 0.15 \pm 0.08$ and $T=0.27 \pm 0.06$ with $U=0$.  
In the framework of the two-Higgs-doublet model, 
we have studied the effects of the CDF $m_W$ measurement together with other
constraints, which include theoretical requirements (potential bounded
from below, unitarity, perturbativity, vacuum stability),
flavor-changing neutral currents in $B$ physics, the cutoff scale above $1\tev$,
Higgs precision data, and direct collider search limits from the
LEP, Tevatron, and LHC.
Pursuing the comprehensive and comparative study,
we consider 16 cases, type I,  type II,  type X, and  type Y
for the normal and inverted Higgs scenarios before and after the CDF $m_W$ measurement.
The still-valid parameter space has been illustrated.
The most unprecedented consequence is the upper bounds on
the masses of the heavy Higgs boson
$M_{H, A, H^\pm} \lsim 1.1\tev $  in the normal scenario and
$M_{H^\pm (A)} \lsim 450\,(420)\gev $ in the inverted scenario. 
Such interesting findings imply that the upcoming LHC run can readily
close out a significant portion of the still-available parameter space.

Before closing, a few more findings from our study are offered as follows:
\begin{enumerate}
 \item The updated fit on the oblique parameters of $S$ and $T$ indicates that the new $m_W$
  measurement requires larger mass splittings among the isospin components
  in multiplet models.

  \item 
   In subsequent steps of imposing constraints on the parameters, we found that
  the survival percentages in the CDF case are much
  smaller than those in the PDG case, thus implying more restriction on physics
  beyond the SM (not only for 2HDM).  This behavior has been
  demonstrated in all four types and two Higgs scenarios of the 2HDM.

\item The $\tan\beta$ is bounded from above more severely when 
  the new $m_W^{\rm CDF}$ value is used in the normal scenario:
  $\tan\beta \lsim 45,\, 8,\, 43,\, 17$ for type I, II, X, and Y, respectively.
  On the other hand, $\tan\beta$ is not bounded in type I and
  type X in the inverted scenario.  There is no parameter space satisfying all the
  requirements in type II and type Y of the inverted scenario.

\item If we raise the cutoff scale $\Lambda_c$ beyond 1 TeV, the restriction on
  the parameter space would become more severe.

\end{enumerate}

\acknowledgments
We would like to thank Dr.\,Jin-Hwan Cho of \textit{National Institute for Mathematical Sciences} in Korea
for helping the numerical computation.
K.C. was supported by MoST with grant numbers MOST-110-2112-M-017-MY3.
The work of JK, SL, and JS is supported by 
the National Research Foundation of Korea, Grant No.~NRF-2022R1A2C1007583. 


\end{document}